\documentclass[twocolumn,aps, prl, superscriptaddress, reprint, twocolumn, notitlepage, showpacs]{revtex4-1}
\usepackage[latin9]{inputenc}
\usepackage{verbatim}
\usepackage{amsmath}
\usepackage{amsthm}
\usepackage{amssymb}
\usepackage{graphicx}
\usepackage[colorlinks=true,linkcolor=blue,citecolor=blue,urlcolor=blue]{hyperref}

\makeatletter
\usepackage{bbding}
\usepackage{xcolor}
\usepackage{amsthm}\usepackage{amsfonts}\usepackage{enumitem}\usepackage{dsfont}
\usepackage{yfonts}\usepackage{calligra}
\usepackage{graphicx}
\usepackage{pxfonts}
\usepackage{upgreek}%\usepackage{lmodern}
\usepackage{geometry}

\providecommand{\U}[1]{\protect\rule{.1in}{.1in}}

\makeatother

\begin{document}
\title{Binary Homodyne Detection \\ for Observing Quadrature Squeezing in Satellite Links}

\author{Christian R. M\"{u}ller}
\affiliation{Max Planck Institute for the Science of Light, Staudtstr. 2, 91058 Erlangen, Germany}
\affiliation{Institute of Optics, Information and Photonics, University of Erlangen-Nuremberg, Staudtstr. 7/B2, D-91058 Erlangen, Germany}

\author{Kaushik P. Seshadreesan}
\affiliation{Max Planck Institute for the Science of Light, Staudtstr. 2, 91058 Erlangen, Germany}
\affiliation{Institute of Optics, Information and Photonics, University of Erlangen-Nuremberg, Staudtstr. 7/B2, D-91058 Erlangen, Germany}

\author{Christian Peuntinger}
\affiliation{Max Planck Institute for the Science of Light, Staudtstr. 2, 91058 Erlangen, Germany}
\affiliation{Institute of Optics, Information and Photonics, University of Erlangen-Nuremberg, Staudtstr. 7/B2, D-91058 Erlangen, Germany}

\author{Christoph Marquardt}
\affiliation{Max Planck Institute for the Science of Light, Staudtstr. 2, 91058 Erlangen, Germany}
\affiliation{Institute of Optics, Information and Photonics, University of Erlangen-Nuremberg, Staudtstr. 7/B2, D-91058 Erlangen, Germany}

\email{christian.mueller@mpl.mpg.de}

\date{\today}

\begin{abstract}
Optical satellite links open up new prospects for realizing quantum physical 
experiments over unprecedented length scales. 
We analyze and affirm the feasibility of detecting quantum squeezing in 
an optical mode with homodyne detection of only one bit resolution, 
as is found in satellites already in orbit. 
We show experimentally that, in combination with a coherent displacement, 
a binary homodyne detector can still detect quantum squeezing efficiently 
even under high loss. 
The sample overhead in comparison to non-discretized homodyne 
detection is merely a factor of 3.3.

\end{abstract}
\maketitle

The laws of quantum mechanics have been validated by numerous 
fundamental tests \cite{ShadboltReview2014}.
With the advent of optical satellite links 
\cite{Rarity2002,Ursin2009, Merali2012,GKESB16, Vedovato2017} 
it is now possible to also validate quantum mechanics over vast distances and a varying gravitational potentials. 
This includes non-classical states \cite{Rideout2012} such 
as quadrature squeezed states of the light field \cite{WPGCRSL12, AGML16}. 
Squeezing is efficiently measured via homodyne detection \cite{YC83}, 
a measurement technique of utmost importance not only in optics, but 
in diverse physical architectures such as 
optomechanical resonators~\cite{Marquardt2014}, 
superconducting qubits~\cite{Schoelkopf2007, Wallraff2009}, 
spin ensembles~\cite{KP03, CLP07, NTTT11, MMKS16} and 
Bose-Einstein condensates~\cite{Oberthaler2008}.
Homodyne detection yields continuously distributed quadrature projections, 
which in practice are sometimes deliberately discretized. 
In optical quantum information processing ~\cite{BvL05, SBRF93, Leo97, LR09, MPDKV16} 
this is exemplified by quantum key distribution protocols \cite{ZHRL09}, and 
by tests of Bell's inequalities~\cite{GDR99, MCVBell99, AUBERSON2002327, BW99, SWKLD16}, 
which inherently require to discretize the homodyne outcomes to binary values.

Optical homodyne detectors are ubiquitous in telecommunications 
and can even be found on optical satellites already in orbit.
Such satellites are promising candidates for exploring quantum technology 
and bringing fundamental tests of quantum mechanics to space both 
rapidly and cost-effectively.
Satellite links, however, imply considerable channel loss which 
reduces the observable squeezing value. 
Moreover, as currently the primary application of optical satellites is 
classical communication via binary phase-shift keying, 
only the sign of the homodyne signal is relevant and the data 
is often projected into binary outcomes during signal processing \cite{SGR16, HMZ15}. 
The question arises whether under such strong technical constraints 
quadrature squeezing can still be detected. 

Extreme discretization into binary outcomes has been studied 
extensively for photon number measurements. 
Photon ``on-off" detection 
and the photon number parity measurement 
were shown to allow for (near-) optimal applications in 
quantum state discrimination \cite{Do73, WTCSLA08, TS08, WATSL10, WGTL12} and 
quantum optical metrology~\cite{DS15, DJK15, SKDL13, SALD11, ARCPH10, MORDOR15}. 
Discretized homodyne detection schemes were used for witnessing 
single photon entanglement~\cite{MBHSD13} 
and for super-resolved imaging with coherent states~\cite{DJA13}.

In this Letter, we investigate fundamental limits of discretized homodyne measurements, 
particularly focusing on the detection of quadrature squeezing. 
We consider the extreme case of a binary homodyne detector (BHD) that 
simply distinguishes between positive and negative quadrature values 
and we analyze its performance for the detection 
of individual signals as well as for the consecutive detection 
of multiple copies of the same state. 
We show that despite the extreme constraint BHD can detect quadrature 
squeezing efficiently even under unfavorable conditions like high loss, when relying on ensemble measurements. 
The ratio between the required number of copies to obtain 
the same information about the observed signal - measured in terms 
of the Bayesian \textit{a posteriori} probability - is merely 3.3 and 
is independent on the squeezing parameter. 
We complement our theoretical analysis with an experimental verification. 
To this end, we prepare and detect both a coherent state and a weakly 
squeezed state via BHD detection and compare the results to ideal, 
i.e. non-discretized, homodyne detection.
We finally discuss the feasibility of detecting squeezed states 
via BHD detection in satellite links.

\begin{figure}%
\includegraphics[width=0.95\columnwidth]{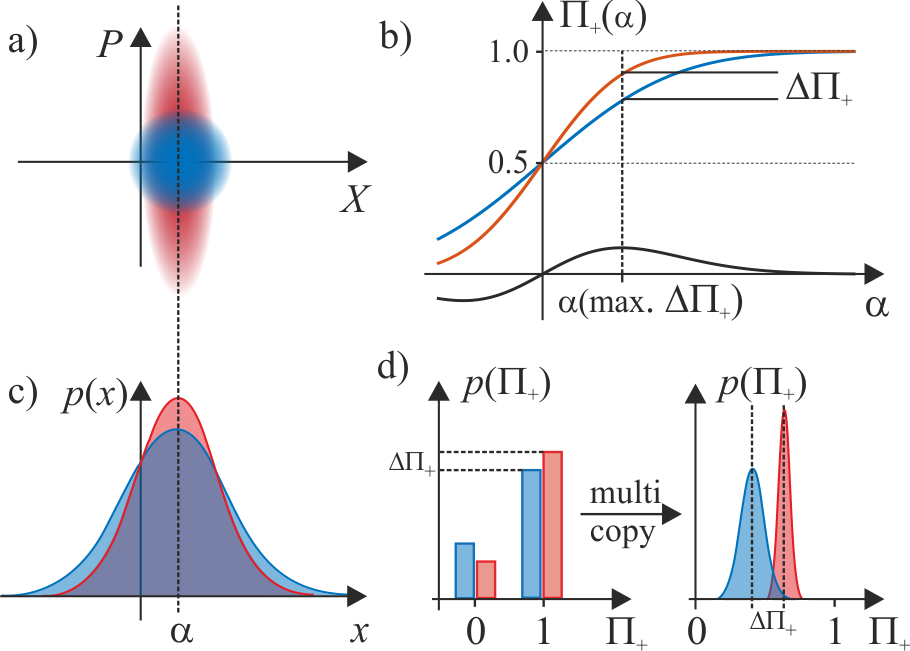}%
\caption{
a) Illustration of a coherent (blue) and a displaced squeezed vacuum state (red) 
in phase space.
b) Binary homodyne expectation value $\Pi_{+}$ of the displaced states and their 
difference $\Delta\Pi_{+}(\alpha)$ (black) as a function of the displacement amplitude $\alpha$.
c) Ideal homodyne marginal distributions along the squeezed $x$-quadrature.
d) Sketch of the statistical BHD probability distribution $p(\Pi_{+})$ 
obtained via single-copy and a multi-copy detection.}%
\label{WitnessingSchemeViaQParity}%
\end{figure}

\noindent\textit{Binary Homodyne Detection--} 
In the following, we introduce the binary homodyne observable and
describe how its expectation value can be controlled via a coherent displacement. 
We consider detecting a Gaussian state, for which the ellipticity \cite{MPDKV16}, i.e. the ratio between the major- and minor semi-axis of its phase space distribution, 
is unknown but has one of two possible values. 
The described detection scheme is insusceptible to whether the state is pure or 
mixed such that we can assume pure states. 
Moreover, the overlap between any two states remains unchanged when applying 
the same squeezing operation to both. 
We therefore assume that one state is a coherent state (the most classical state as expected under the influence of high loss) and 
the other state is a squeezed state with the same mean amplitude 
and with a real-valued squeezing parameter $r$ (see Fig.~\ref{WitnessingSchemeViaQParity}(a)). 

To identify the received signal via a BHD, the signal is first displaced 
in phase space, followed by a projection onto quadrature semi-axes. 
This measurement is described by two positive operator-valued measure elements (POVM)
\begin{eqnarray}
\hat{\Pi}_{+}\left(\alpha\right) &=&\intop_{0}^{\infty}dx\,\hat{D}^{\dagger}\left(\alpha\right)\left|x\right\rangle \left\langle x\right|\hat{D}\left(\alpha\right),\nonumber \\
\hat{\Pi}_{-}\left(\alpha\right) &=& I-\hat{\Pi}_{+}\left(\alpha\right),
\end{eqnarray}
where $\alpha$ is the displacement amplitude.
Owing to the symmetry along the quadrature axis, we can restrict the analysis 
to real and positive displacement amplitudes.

%%%%%%%%%%%%%%%%%%%%%%%%%%%%
%%%%%%%%%%%%%%%%%%%%%%%%%%%%

\vspace{0.25cm}
\noindent \textit{Decision Rule.--} 
The BHD outcome is a Bernoulli random variable $Y(\alpha)$ over the sample space 
$y\in\left\{+,-\right\}$. 
The likelihood functions for the two hypotheses 
$\hat{\rho}_{h}\in\left\{ \mathrm{coh}, \mathrm{sqz}\right\}$ 
are the conditional probabilities 
$P_{Y\left|H\right.}\left(y\left|\,h\right.\right)=\mathrm{Tr}\left[\hat{\Pi}_y\,\hat{\rho}_h \right]$. 
%The likelihood for positive '+' and negative '-' quadrature projections are given by
\begin{eqnarray}
P_{Y\left|H\right.}\left( {+ \atop -} \left|\,\mathrm{coh}, \alpha\right.\right) &=& \frac{1}{2}\left(1\pm\text{erf}\left(\frac{\alpha}{\sqrt{2}}\right)\right),\nonumber \\
P_{Y\left|H\right.}\left( {+ \atop -} \left|\,\mathrm{sqz}, \alpha,r\right.\right) &=& \frac{1}{2}\left(1\pm\text{erf}\left(\frac{\alpha}{\sqrt{2\,e^{-2\,r}}}\right)\right).
\label{equ:ConditionalProbabilityGaussian}
\end{eqnarray}

The average \textit{a posteriori} probability is derived by updating 
the priors ($P_{H}(\mathrm{coh})=P_{H}(\mathrm{sqz})=1/2$) 
via Bayesian inference
% and averaging over the two possible states. 
\begin{equation}
\langle P_{H\left|Y\right.}(\alpha, r) \rangle = \sum_{y = \{+,-\}} \frac{\sum_{h} P_{Y\left|H\right.}(y\,|\,h, \alpha, r)^2}{2\,P_{Y}(y)},
\label{eq:sgl_apos}
\end{equation}
where $P_{Y}(y) = \left( P_{Y\left|H\right.}(y\,|\,\mathrm{coh}) + P_{Y\left|H\right.}(y\,|\,\mathrm{sqz})\right)/2$, and is optimized over the displacement $\alpha$. 

\begin{figure}%
\includegraphics[width=.8 \columnwidth]{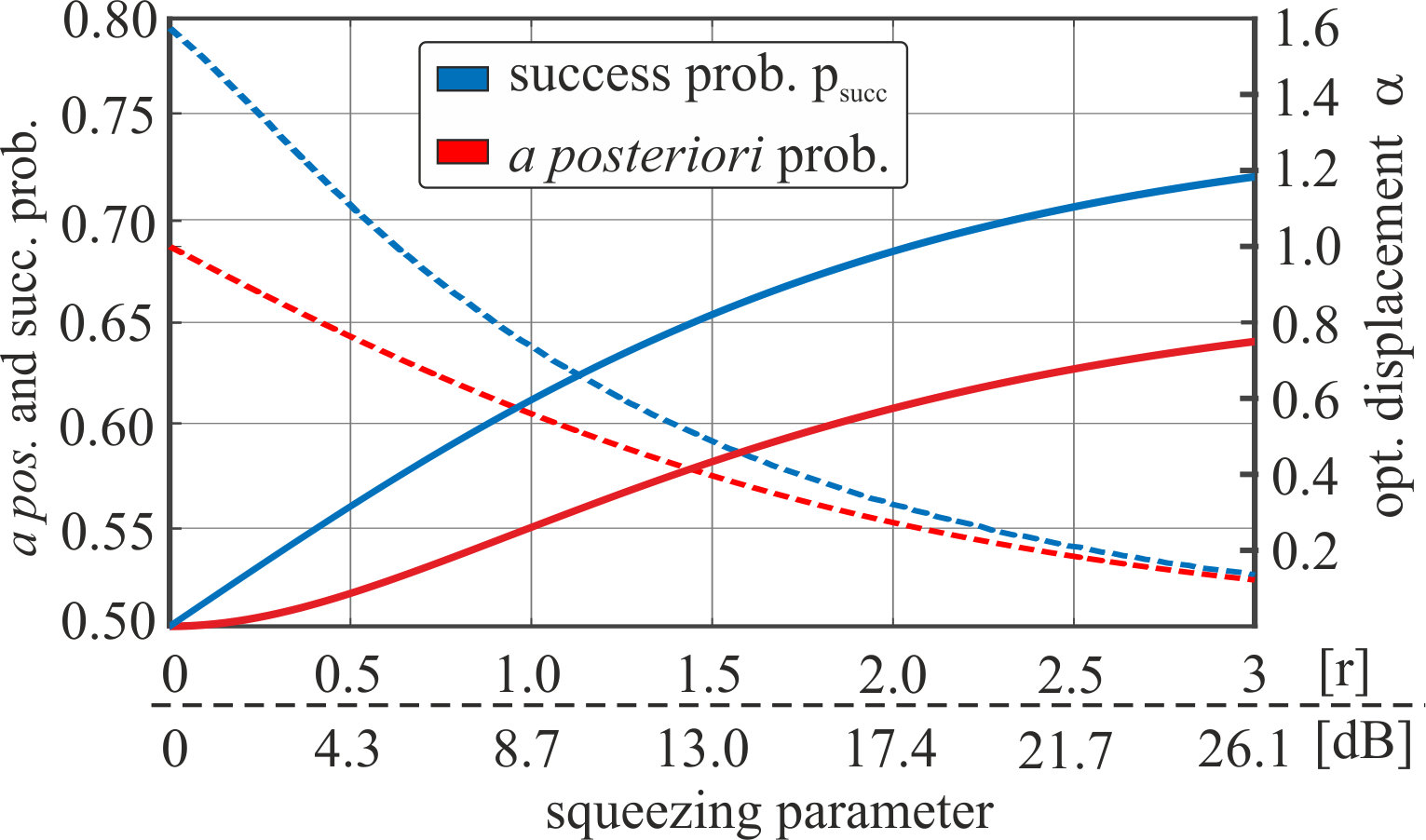}%
\caption{Average \textit{a posteriori} (red) 
and success probability (blue) for the detection 
of a single state, as well as the associated 
optimized displacement amplitudes 
$\alpha$ (dashed lines and right axis labels). }%
\label{apos_single-copy}%
\end{figure}

For positive $\alpha$, 
$P_{Y\left|H\right.}\left(+\left|\,\mathrm{sqz}\right.\right) > P_{Y\left|H\right.}\left(+\left|\,\mathrm{coh}\right.\right)$, 
such that the squeezed state can be associated with the outcome \mbox{'+'} 
and the coherent state with the outcome \mbox{'-'}. 
Fig.~\ref{WitnessingSchemeViaQParity}(b) shows the BHD expectation value 
$\left\langle\Pi_{+}(\alpha)\right\rangle$. % as a function of $\alpha$. 
Varying the displacement amplitude allows maximizing the difference 
$\Delta\Pi_{+}=P_{Y\left|H\right.}\left(+\,|\,\mathrm{sqz}\right) - P_{Y\left|H\right.}\left(+\,|\,\mathrm{coh}\right)$ 
between the expectation values %of the squeezed state and the coherent state 
and consequently allows for an optimized discrimination. 
The maximal success probability for a single detection is 
\begin{equation}
p_{succ}=\max_{\alpha} \frac{1+\Pi_{+}(\alpha)}{2}
\label{eq:avg_err}
\end{equation}

The achievable success- and \textit{a posteriori} 
probability, as well as the associated optimized displacement 
amplitudes are shown in Fig.\ref{apos_single-copy} as a function 
of the squeezing parameter.
Note, that the displacement amplitudes optimizing the two 
parameters coincide only in the limit of large squeezing amplitudes. 
This emphasizes that \textit{a posteriori}- and success probability 
and are indeed distinct figures of merit.
The \textit{a posteriori} probability is maximized by optimizing the 
difference between the conditional probabilities for any possible outcome, 
while the optimal success probability requires maximizing $\Delta\Pi_{+}$.
The optimal displacement in the latter case coincides with the 
intersection point of the states' marginal distributions 
as depicted in Fig.~\ref{WitnessingSchemeViaQParity}(c). 
%For further details see Supplemental Material \ref{optdissingle}.
\begin{equation} 
\alpha^{(opt)}(r) = \sqrt{\frac{2r}{e^{2r}-1}}.
\label{eq:OptimizedDisplacement}
\end{equation}

In the limit of an infinitely squeezed state the optimized 
displacement asymptotically approaches zero 
$\lim_{r\to\infty} \alpha(r)=0$, 
but the success probability is upper-bounded by 
$p_{succ}\leq\frac{3}{4}$, 
as at least half of the coherent state has support on the positive semi-axis.
Similarly, the \textit{a posteriori} probability is upper bounded by 
$\langle P_{H\left|Y\right.}(\alpha, r) \rangle \leq \frac{2}{3}$.

\vspace{0.25cm}
\noindent \textit{Multi-Copy Detection--} 
To verify the properties of the quantum states after propagation it is sufficient to perform ensemble measurements. 
Gathering statistics over multiple measurements of identically prepared states 
allows reducing the overlap of the signals probability distributions 
(see Fig.~\ref{WitnessingSchemeViaQParity}(d) ) 
and consequently reducing the error probability. 
Let $\overrightarrow{y}=\left(y_{1},y_{2},\cdots,y_{N}\right)$
denote the outcome of a multi-copy BHD measurement.

The probability to detect $k \in \{0,1,...N\}$, positive 
quadrature projections from $N$ measurements is given by 
the Binomial probability density function
\begin{equation}
P_{Y\left|H\right.}^{(N)}(k|h,\alpha,r)=\binom{N}{k}\,P_{Y\left|H\right.}(+\left|h,\alpha,r\right.)^{k}\,\left(1-P_{Y\left|H\right.}(+\left|h,\alpha,r\right.)\right)^{N-k}
\label{eq:bino_dist}
\end{equation} 
which approaches a quasi-continuous Gaussian distribution for a large number of samples.

The \textit{a posteriori} probability for the signal hypothesis 
$h\in \left\{\mathrm{coh},\mathrm{sqz}\right\}$ 
follows from the conditional single-copy probabilities in 
Eq.(\ref{equ:ConditionalProbabilityGaussian}) and Eq.(\ref{eq:sgl_apos}) 
via Bayesian inference as
\begin{equation}
P^{(N)}_{H\left|\overrightarrow{Y}\right.}\left( h\left|\overrightarrow{y}\right.\right) = \frac{\prod_{i=1}^{N}P_{Y\left|H\right.}\left(y_{i}\left|h\right.\right)}{\sum_{h}\prod_{j=1}^{N} P_{Y\left|H\right.}\left(y_{j}\left|h\right.\right)}.
\label{eq:MultiCopyAPos}
\end{equation}

\begin{figure}[tb]
\includegraphics[width=.95\columnwidth]{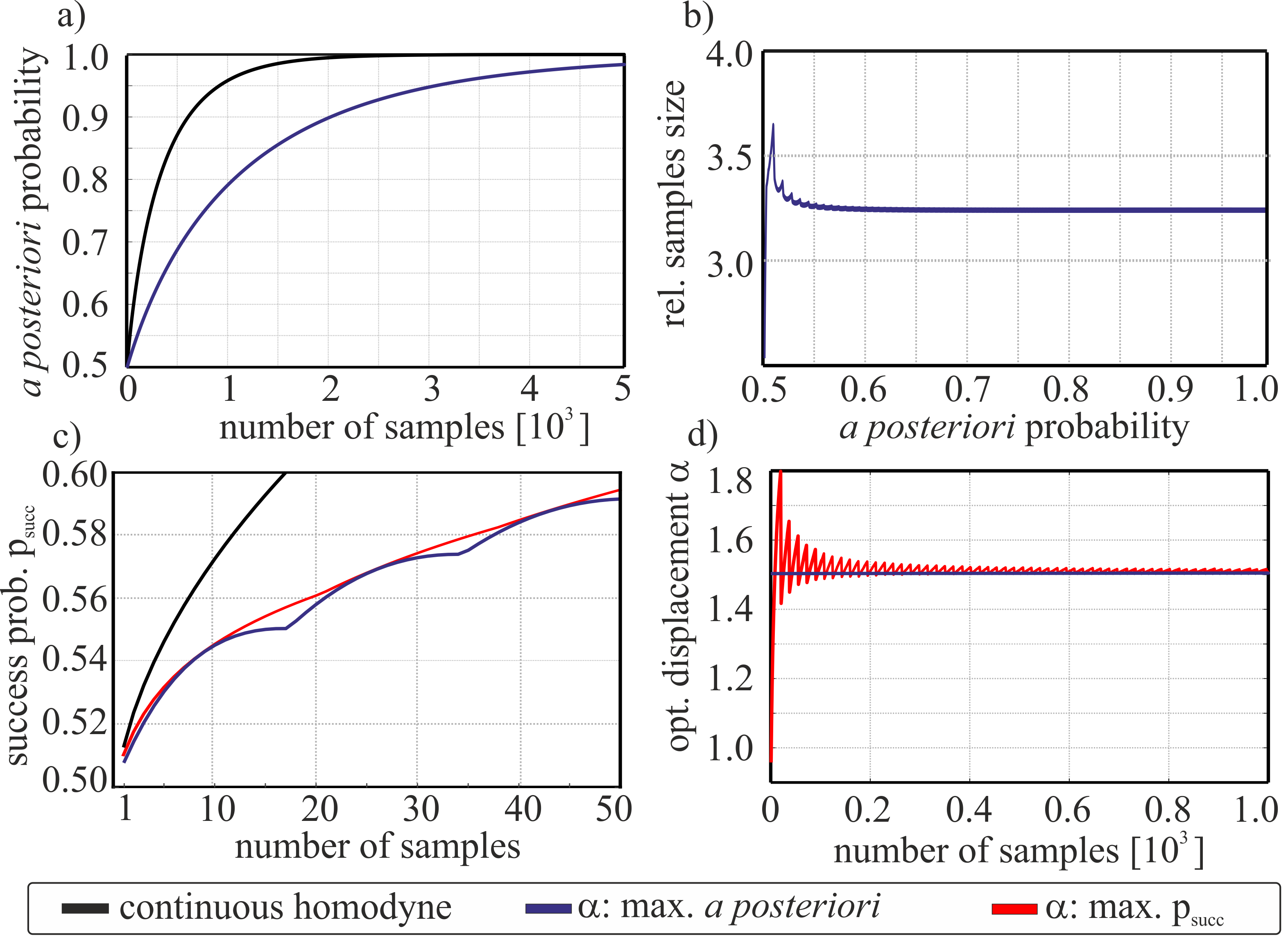}%
\caption{
a) \textit{A posteriori} probabilities for 
non-discretized homodyne detection (black), and BHD detection as a function of the 
number of samples .The displacement is optimized for maximizing the \textit{a posteriori} probability.
b) Relative sample overhead required to obtain the same \textit{a posteriori} probability
for the BHD measurement as obtained by non-discretized homodyne detection.  
c) Probability for successful state discrimination as a function 
of the number of samples and for different displacement strategies. 
d) Comparison of the optimized displacement amplitudes $\alpha$ for different strategies.}
\label{fig:MultiCopyCombined}
\end{figure}

The maximal \textit{a posteriori} probability is obtained by maximizing over the 
displacement parameter $\alpha$ and is given by 
\begin{equation}
\langle P^{(N)}_{H\left|\overrightarrow{Y}\right.}\rangle = \max_{\alpha} \sum_{k=0}^{N} \frac{\sum_{h}\left[P_{Y\left|H\right.}^{(N)}(k\,|\,h,\alpha, r)\right]^2 }{2 \sum_{h'} P_{Y\left|H\right.}^{(N)}(k\,|\,h',\alpha, r)}.
\label{eq:aposteriori}
\end{equation}

This quantity is compared to an non-discretized homodyne detector 
in Fig.~\ref{fig:MultiCopyCombined}(a) for the example of a 
squeezing parameter $r=0.085$, i.e. 0.369\,dB below the shot noise level.
Naturally, the outcomes of the BHD yield less information 
about the detected states than ideal homodyne detection. 
(For details on the calculation of the \textit{a posteriori}- and success 
probability for ideal homodyne detection see Supplemental Material \ref{aposHD}.)
Consequently, a larger number of samples has to be collected on average 
to achieve the same \textit{a posteriori} probability. 
The required relative sample size is depicted in Fig.~\ref{fig:MultiCopyCombined}(b). 
Apart from the region of \textit{a posteriori} probabilities 
close to $0.5$, where statistical effects originating from the 
discreteness of the Binomial distribution are most pronounced, 
the ratio is essentially constant and has a value of merely 3.3.
A separate analysis shows that this ratio is independent of 
the squeezing parameter $r$.
We stress that this value, as it applies to the extreme 
discretization into binary outcomes, is indeed an upper 
bound for the required sample overhead of arbitrarily 
discretized homodyne detection as e.g. occurring in realistic 
analog-to-digital (AD) converters.

The optimal success probability $p_{succ}$ in multi-copy detection
is obtained by generalizing Eq.(\ref{eq:avg_err}) such that the squeezed 
state is hypothesized whenever $k$ equals or exceeds 
a threshold value $\tau$, followed by maximizing over $\tau$ and $\alpha$.
\begin{equation}
P^{(N)}_{\mathrm{succ}}\left(k\right) = \max_{\alpha,\tau} \frac{P^{(N)}(k\leq\tau|\,\mathrm{coh},\alpha)+P^{(N)}(k>\tau|\,\mathrm{sqz},\alpha,r))}{2}
\label{eq:erro_prob}
\end{equation}
Fig.~\ref{fig:MultiCopyCombined}(c) compares the optimized success probability 
of the BHD to an ideal homodyne detector as well as to a BHD where the
displacement maximizing the \textit{a posteriori} probability was applied. 
Maximizing the \textit{a posteriori} probability is only equivalent 
to minimizing the error probability in the limit of a large number of 
samples.
This is underpinned by Fig.~\ref{fig:MultiCopyCombined}(d) where the 
distinct optimized displacements are illustrated.
The fluctuations in the displacement maximizing the success
probability originate from the discreteness of the underlying 
Binomial distribution. 
A more detailed discussion
%on the example of the first kink between 20 and 21 samples 
is given in Supplemental Material Sec.\ref{Discontinuity}.

\begin{figure}
  \includegraphics[width=\columnwidth]{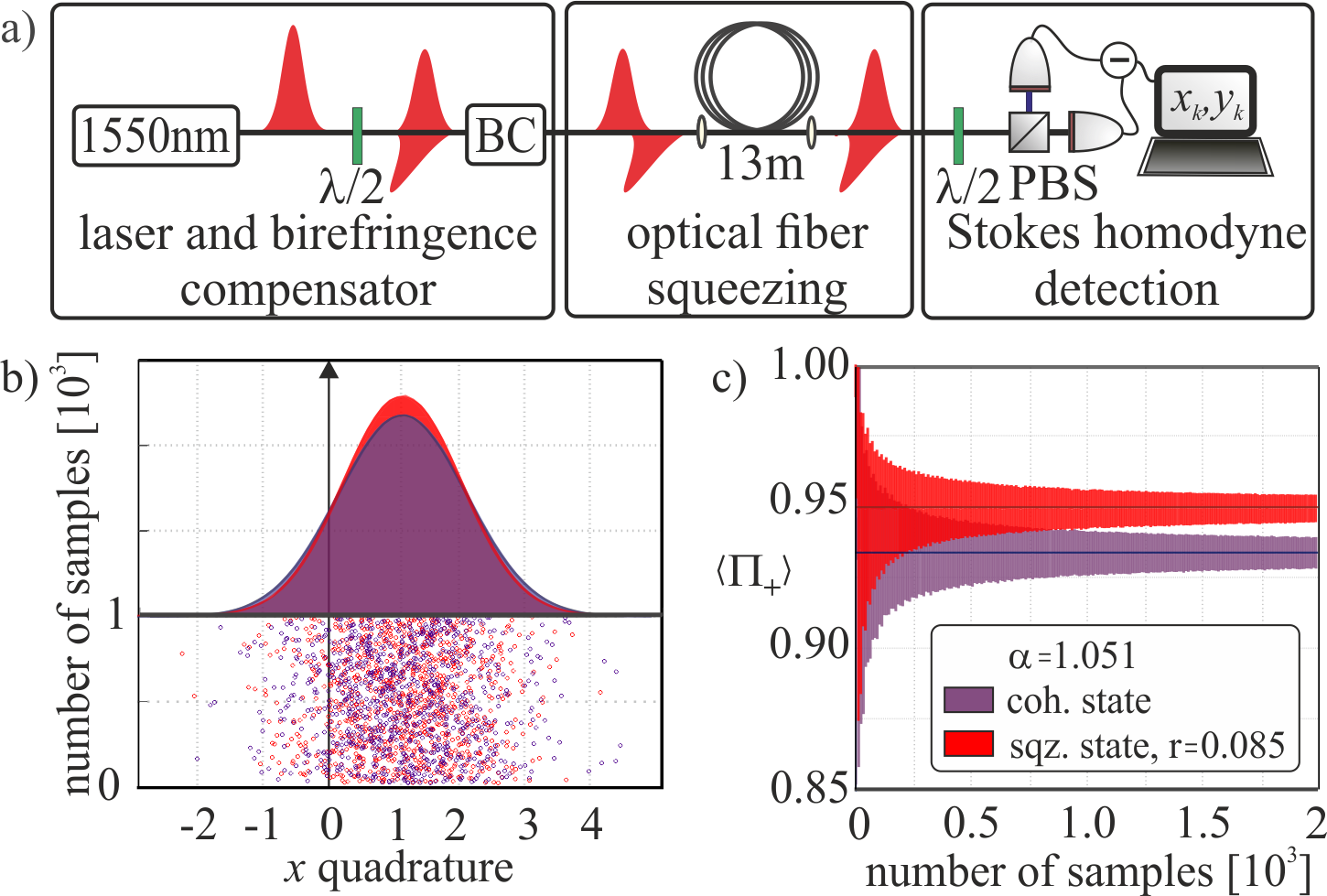}
  \caption{
	a) Outline of the experimental setup. 
	b) Measured marginal distributions of the coherent state (purple) and of 
	   the squeezed state (red) with $r=0.085$; i.e. 0.369\,dB.
	   An excerpt of displaced homodyne samples for both states is displayed in the lower section.
	c) Statistical distribution of the quadrature-parity observable 
	   as a function of the observed number of samples. 
		 The solid lines indicate the average quadrature-parity.}
  \label{fig:setup}
\end{figure}

\vspace{0.25cm}
\noindent \textit{Experimental Validation.--}
We experimentally prepared a vacuum state and a weakly squeezed vacuum 
state ($r=0.085$; 0.369\,dB) using the fiber-based polarization squeezing 
setup~\cite{HJLA2005} outlined in  Fig.~\ref{fig:setup}(a).
%%%%%%%%%%%%%%%%%%%%%%%%%%%%
%%%%%%%%%%%%%%%%%%%%%%%%%%
Pulses from a shot-noise limited laser centered at 1560\,nm 
are distributed equally on the principal axes of a 13\,m long 
polarization maintaining fiber and get individually 
squeezed via the Kerr nonlinearity \cite{KY86}. 
The emerging pulses are locked to circular polarization 
such that the mean value of the squeezed polarization state lies 
along the $S_3$ direction on the Poincar\'{e} sphere. 
Homodyne detection of the quantum Stokes variables within 
the \textit{dark} $S_1-S_2$ plane is then equivalent 
to conventional homodyne detection in the canonical 
\textit{x-p} phase space. A detailed description of the 
setup and the theoretical background can be found in \cite{MSPG12}. 
In total, we acquired $2 \cdot10^8$ homodyne samples $\{x_k\}$ 
of identically prepared copies of the vacuum and the squeezed state 
projected along the squeezed quadrature.
The detector resolution is \mbox{16 bit}, such that the quadrature was 
sampled quasi-continuously and the data provides accurate histograms 
of the actual marginal distribution as shown in Fig.~\ref{fig:setup}(b).
Applying a coherent displacement solely adds a \textit{classical} 
offset amplitude to the quantum field operator, 
$\hat{a} \mapsto \hat{a}+\alpha$, 
and consequently to the detected quadrature, 
$\hat{X}\mapsto\hat{X}+\mathrm{Re}(\alpha)$, 
where 
$\hat{X} = (\hat{a}+\hat{a}^{\dagger})/2$.
Therefore, given the quasi-continuous homodyne samples $\{x_k\}$, the coherent 
displacement and the subsequent projection onto the quadrature semi-axes $y_k \in \left\{+,-\right\}$
can faithfully be performed after detection via
$x_k \mapsto y_k = \frac{1}{2}\left(\mathrm{sign}\left(x_k + \alpha\right)+1\right)$. \\

The statistical distribution of the BHD samples for the measured 
coherent (purple) and squeezed state (red) are shown in Fig.~\ref{fig:setup}(c) 
for a displacement amplitude $\alpha = 1.501$ which maximizes the \textit{a posteriori} 
probability. 
The distributions are largely overlapping for a small number of samples, but eventually 
become distinguishable with increasing sample size.

%%%%%%%%%%%%%%%%%%%%%%%%%%
\begin{figure}[t]%
\centering
\includegraphics[width=.9\columnwidth]{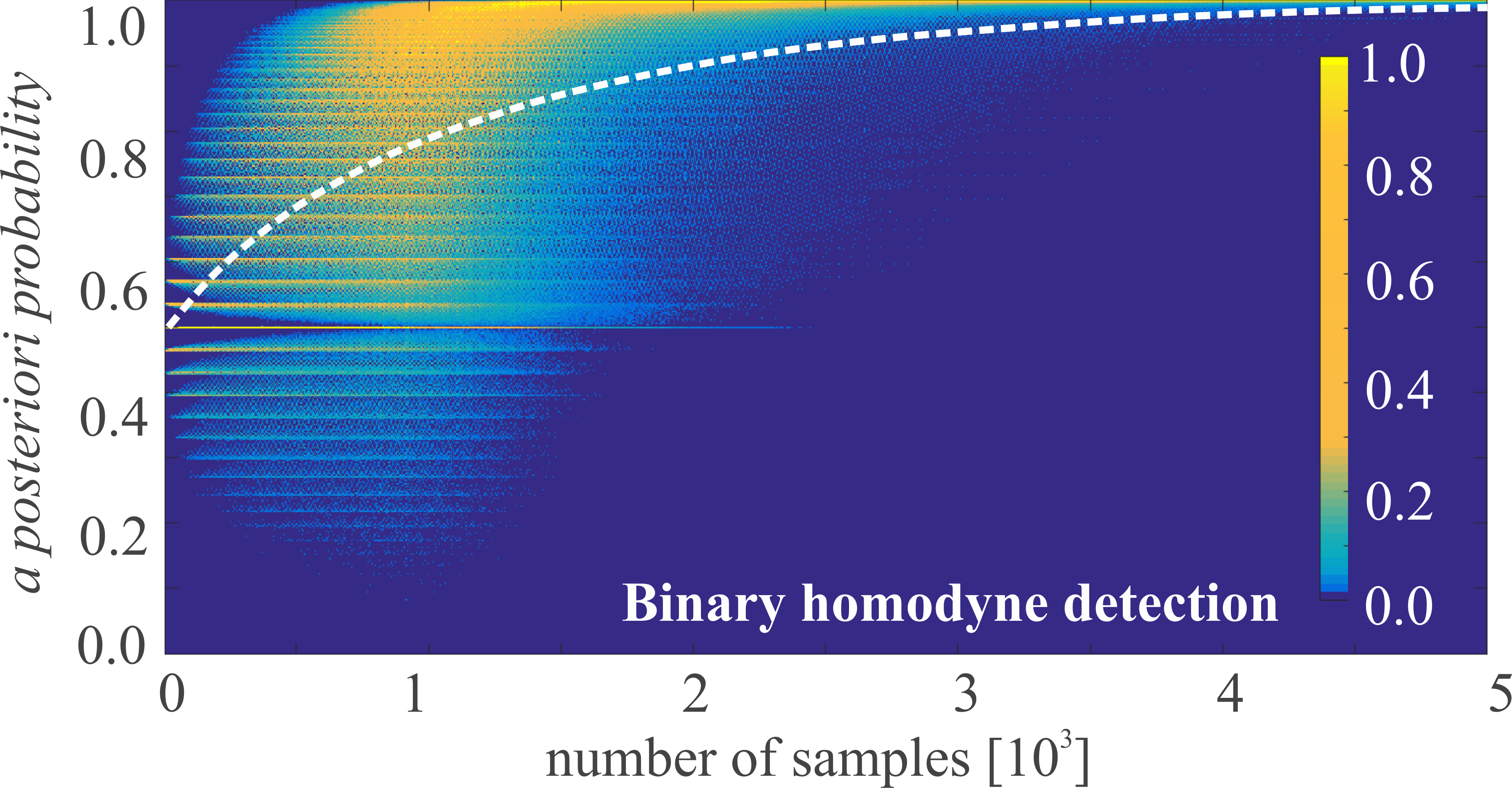}%
\caption{
Statistical distributions of the experimentally obtained 
\textit{a posteriori} probabilities as a function of the 
number of detected samples $N$. 
For improved contrast, the maximal \textit{a posteriori} 
probability for each value of $N$ has been normalized to unity.
The dashed white curve shows the average value for each number of detections.}
\label{fig:StaticParityValueScaling}%
\end{figure}

Fig.~\ref{fig:StaticParityValueScaling} shows the experimentally obtained 
statistical distribution of the \textit{a posteriori} probabilities for BHD 
detection as a function of the number of samples.
The discreteness of the sample distributions yields a rich structure which 
is consistent with numerical simulations.
For a large number of samples the \textit{a posteriori} probability 
converges to unity as could be expected.
The dashed white curve shows the average \textit{a posteriori} probability 
which also is in excellent agreement with theoretical predictions.

\vspace{0.25cm}
\noindent \textit{Observing Squeezing in Optical Satellite Links.--}
Optical communication satellites are currently optimized for 
the detection of binary encoded classical signals. 
This encompasses projecting the signals into binary outcomes 
already in the detection process, which thus prevents 
an evaluation of the continuous-variable quantum information. 
Our results on BHD detection pave the way to nevertheless
detect quadrature squeezing in such scenarios.
Let us investigate how the size of the measurement data required 
to faithfully detect a squeezed state scales as a function of loss 
in the transmission channel 
(see Supplemental Material \ref{datasize} for details). 
An up-link from a ground-based sender to a low Earth orbit (LEO) 
satellite receiver ($\approx600$ km distance) with transmitter 
and receiver apertures of about $30$\,cm diameter is subject to 
transmission losses of about 40-45\,dB~\cite{BMHHE13}. 
These losses includes atmospheric turbulence, diffraction and pointing error. 
Let us further assume a moderate squeezing parameter of $r = 0.69$, 
corresponding to $6$\,dB of squeezing below the shot noise level, 
and let us bound the average error probability to at most $10^{-2}$. 
To satisfy this realistic scenario for 40\,dB of loss, about $3\cdot10^{9}$ 
BHD samples are required, while 45\,dB of loss would require $3\cdot10^{10}$ samples. 
We stress again that this is of the same order of magnitude as the number 
of samples required by a non-discretized homodyne detector, differing 
merely by a constant factor of 3.3. 
Optical homodyne detection and quadrature squeezing have been demonstrated 
with GHz bandwidths~\cite{Gabriel12, Ast2013}, such that the acquisition 
of the required number of samples can be achieved within a few seconds, 
i.e. well within the typical link time for a single flyover of a 
LEO satellite ($\approx300$ seconds). 
We thus emphasize that it is, in principle, feasible with technology already in orbit to 
detect squeezing in optical satellite links. 

\vspace{0.25cm}
\noindent \textit{Conclusions.--}
Homodyne detection is a powerful quantum measurement even under 
the constraint of severe discretization.
Our work underlines this by showing that a homodyne detector with 
a resolution of only one bit can still accomplish the quintessential 
task of detecting squeezed light efficiently. 
We derived the optimal phase space displacements for maximizing the 
success probability and the \textit{a posteriori} probability, 
and found that a binary homodyne detector can distinguish two 
Gaussian states of different squeezing parameter with the same 
\textit{a posteriori} probabilities as its non-discretized counterpart 
by detecting a sample set merely a factor of 3.3 larger.
Our work opens prospects for detecting quantum squeezing 
in optical satellite links. 
A possible early experimental demonstration might involve using 
(binary) homodyne detection on the satellite and an 
Earth-to-satellite communication link with a squeezer at the ground station, that includes the necessary techniques for phase front precompensation and phase locking.\newline

\noindent We thank Gerd Leuchs, Ulrik L. Andersen and Dominique Elser for fruitful discussions.

\bibliography{References}

\begin{thebibliography}{50}%
\makeatletter
\providecommand \@ifxundefined [1]{%
 \@ifx{#1\undefined}
}%
\providecommand \@ifnum [1]{%
 \ifnum #1\expandafter \@firstoftwo
 \else \expandafter \@secondoftwo
 \fi
}%
\providecommand \@ifx [1]{%
 \ifx #1\expandafter \@firstoftwo
 \else \expandafter \@secondoftwo
 \fi
}%
\providecommand \natexlab [1]{#1}%
\providecommand \enquote  [1]{``#1''}%
\providecommand \bibnamefont  [1]{#1}%
\providecommand \bibfnamefont [1]{#1}%
\providecommand \citenamefont [1]{#1}%
\providecommand \href@noop [0]{\@secondoftwo}%
\providecommand \href [0]{\begingroup \@sanitize@url \@href}%
\providecommand \@href[1]{\@@startlink{#1}\@@href}%
\providecommand \@@href[1]{\endgroup#1\@@endlink}%
\providecommand \@sanitize@url [0]{\catcode `\\12\catcode `\$12\catcode
  `\&12\catcode `\#12\catcode `\^12\catcode `\_12\catcode `\%12\relax}%
\providecommand \@@startlink[1]{}%
\providecommand \@@endlink[0]{}%
\providecommand \url  [0]{\begingroup\@sanitize@url \@url }%
\providecommand \@url [1]{\endgroup\@href {#1}{\urlprefix }}%
\providecommand \urlprefix  [0]{URL }%
\providecommand \Eprint [0]{\href }%
\providecommand \doibase [0]{http://dx.doi.org/}%
\providecommand \selectlanguage [0]{\@gobble}%
\providecommand \bibinfo  [0]{\@secondoftwo}%
\providecommand \bibfield  [0]{\@secondoftwo}%
\providecommand \translation [1]{[#1]}%
\providecommand \BibitemOpen [0]{}%
\providecommand \bibitemStop [0]{}%
\providecommand \bibitemNoStop [0]{.\EOS\space}%
\providecommand \EOS [0]{\spacefactor3000\relax}%
\providecommand \BibitemShut  [1]{\csname bibitem#1\endcsname}%
\let\auto@bib@innerbib\@empty
%</preamble>
\bibitem [{\citenamefont {Shadbolt}\ \emph {et~al.}(2014)\citenamefont
  {Shadbolt}, \citenamefont {Mathews}, \citenamefont {Laing},\ and\
  \citenamefont {O'Brien}}]{ShadboltReview2014}%
  \BibitemOpen
  \bibfield  {author} {\bibinfo {author} {\bibfnamefont {P.}~\bibnamefont
  {Shadbolt}}, \bibinfo {author} {\bibfnamefont {J.~C.~F.}\ \bibnamefont
  {Mathews}}, \bibinfo {author} {\bibfnamefont {A.}~\bibnamefont {Laing}}, \
  and\ \bibinfo {author} {\bibfnamefont {J.~L.}\ \bibnamefont {O'Brien}},\
  }\href {\doibase 10.1038/nphys2931} {\bibfield  {journal} {\bibinfo
  {journal} {Nature Physics}\ }\textbf {\bibinfo {volume} {10}},\ \bibinfo
  {pages} {278} (\bibinfo {year} {2014})}\BibitemShut {NoStop}%
\bibitem [{\citenamefont {Rarity}\ \emph {et~al.}(2002)\citenamefont {Rarity},
  \citenamefont {Tapster}, \citenamefont {Gorman},\ and\ \citenamefont
  {Knight}}]{Rarity2002}%
  \BibitemOpen
  \bibfield  {author} {\bibinfo {author} {\bibfnamefont {J.~G.}\ \bibnamefont
  {Rarity}}, \bibinfo {author} {\bibfnamefont {P.~R.}\ \bibnamefont {Tapster}},
  \bibinfo {author} {\bibfnamefont {P.~M.}\ \bibnamefont {Gorman}}, \ and\
  \bibinfo {author} {\bibfnamefont {P.}~\bibnamefont {Knight}},\ }\href
  {http://stacks.iop.org/1367-2630/4/i=1/a=382} {\bibfield  {journal} {\bibinfo
   {journal} {New Journal of Physics}\ }\textbf {\bibinfo {volume} {4}},\
  \bibinfo {pages} {82} (\bibinfo {year} {2002})}\BibitemShut {NoStop}%
\bibitem [{\citenamefont {{Ursin, R.}}\ \emph {et~al.}(2009)\citenamefont
  {{Ursin, R.}}, \citenamefont {{Jennewein, T.}}, \citenamefont {{Kofler, J.}},
  \citenamefont {{Perdigues, J. M.}}, \citenamefont {{Cacciapuoti, L.}},
  \citenamefont {{de Matos, C. J.}}, \citenamefont {{Aspelmeyer, M.}},
  \citenamefont {{Valencia, A.}}, \citenamefont {{Scheidl, T.}}, \citenamefont
  {{Acin, A.}}, \citenamefont {{Barbieri, C.}}, \citenamefont {{Bianco, G.}},
  \citenamefont {{Brukner, C.}}, \citenamefont {{Capmany, J.}}, \citenamefont
  {{Cova, S.}}, \citenamefont {{Giggenbach, D.}}, \citenamefont {{Leeb, W.}},
  \citenamefont {{Hadfield, R. H.}}, \citenamefont {{Laflamme, R.}},
  \citenamefont {{L\"{u}tkenhaus, N.}}, \citenamefont {{Milburn, G.}},
  \citenamefont {{Peev, M.}}, \citenamefont {{Ralph, T.}}, \citenamefont
  {{Rarity, J.}}, \citenamefont {{Renner, R.}}, \citenamefont {{Samain, E.}},
  \citenamefont {{Solomos, N.}}, \citenamefont {{Tittel, W.}}, \citenamefont
  {{Torres, J. P.}}, \citenamefont {{Toyoshima, M.}}, \citenamefont
  {{Ortigosa-Blanch, A.}}, \citenamefont {{Pruneri, V.}}, \citenamefont
  {{Villoresi, P.}}, \citenamefont {{Walmsley, I.}}, \citenamefont {{Weihs,
  G.}}, \citenamefont {{Weinfurter, H.}}, \citenamefont {{Zukowski, M.}},\ and\
  \citenamefont {{Zeilinger, A.}}}]{Ursin2009}%
  \BibitemOpen
  \bibfield  {author} {\bibinfo {author} {\bibnamefont {{Ursin, R.}}}, \bibinfo
  {author} {\bibnamefont {{Jennewein, T.}}}, \bibinfo {author} {\bibnamefont
  {{Kofler, J.}}}, \bibinfo {author} {\bibnamefont {{Perdigues, J. M.}}},
  \bibinfo {author} {\bibnamefont {{Cacciapuoti, L.}}}, \bibinfo {author}
  {\bibnamefont {{de Matos, C. J.}}}, \bibinfo {author} {\bibnamefont
  {{Aspelmeyer, M.}}}, \bibinfo {author} {\bibnamefont {{Valencia, A.}}},
  \bibinfo {author} {\bibnamefont {{Scheidl, T.}}}, \bibinfo {author}
  {\bibnamefont {{Acin, A.}}}, \bibinfo {author} {\bibnamefont {{Barbieri,
  C.}}}, \bibinfo {author} {\bibnamefont {{Bianco, G.}}}, \bibinfo {author}
  {\bibnamefont {{Brukner, C.}}}, \bibinfo {author} {\bibnamefont {{Capmany,
  J.}}}, \bibinfo {author} {\bibnamefont {{Cova, S.}}}, \bibinfo {author}
  {\bibnamefont {{Giggenbach, D.}}}, \bibinfo {author} {\bibnamefont {{Leeb,
  W.}}}, \bibinfo {author} {\bibnamefont {{Hadfield, R. H.}}}, \bibinfo
  {author} {\bibnamefont {{Laflamme, R.}}}, \bibinfo {author} {\bibnamefont
  {{L\"{u}tkenhaus, N.}}}, \bibinfo {author} {\bibnamefont {{Milburn, G.}}},
  \bibinfo {author} {\bibnamefont {{Peev, M.}}}, \bibinfo {author}
  {\bibnamefont {{Ralph, T.}}}, \bibinfo {author} {\bibnamefont {{Rarity,
  J.}}}, \bibinfo {author} {\bibnamefont {{Renner, R.}}}, \bibinfo {author}
  {\bibnamefont {{Samain, E.}}}, \bibinfo {author} {\bibnamefont {{Solomos,
  N.}}}, \bibinfo {author} {\bibnamefont {{Tittel, W.}}}, \bibinfo {author}
  {\bibnamefont {{Torres, J. P.}}}, \bibinfo {author} {\bibnamefont
  {{Toyoshima, M.}}}, \bibinfo {author} {\bibnamefont {{Ortigosa-Blanch, A.}}},
  \bibinfo {author} {\bibnamefont {{Pruneri, V.}}}, \bibinfo {author}
  {\bibnamefont {{Villoresi, P.}}}, \bibinfo {author} {\bibnamefont {{Walmsley,
  I.}}}, \bibinfo {author} {\bibnamefont {{Weihs, G.}}}, \bibinfo {author}
  {\bibnamefont {{Weinfurter, H.}}}, \bibinfo {author} {\bibnamefont
  {{Zukowski, M.}}}, \ and\ \bibinfo {author} {\bibnamefont {{Zeilinger,
  A.}}},\ }\href {\doibase 10.1051/epn/2009503} {\bibfield  {journal} {\bibinfo
   {journal} {Europhysics News}\ }\textbf {\bibinfo {volume} {40}},\ \bibinfo
  {pages} {26} (\bibinfo {year} {2009})}\BibitemShut {NoStop}%
\bibitem [{\citenamefont {Merali}(2012)}]{Merali2012}%
  \BibitemOpen
  \bibfield  {author} {\bibinfo {author} {\bibfnamefont {Z.}~\bibnamefont
  {Merali}},\ }\href
  {https://search.proquest.com/docview/1266361041?accountid=105981} {\bibfield
  {journal} {\bibinfo  {journal} {Nature}\ }\textbf {\bibinfo {volume} {492}},\
  \bibinfo {pages} {22} (\bibinfo {year} {2012})}\BibitemShut {NoStop}%
\bibitem [{\citenamefont {G\"{u}nthner}\ \emph {et~al.}(2017)\citenamefont
  {G\"{u}nthner}, \citenamefont {Khan}, \citenamefont {Elser}, \citenamefont
  {Stiller}, \citenamefont {Bayraktar}, \citenamefont {M\"{u}ller},
  \citenamefont {Sauke}, \citenamefont {Tr\"{o}ndle}, \citenamefont {Heine},
  \citenamefont {Seel}, \citenamefont {Greulich}, \citenamefont {Zech},
  \citenamefont {G\"{u}tlich}, \citenamefont {Philipp-May}, \citenamefont
  {Marquardt},\ and\ \citenamefont {Leuchs}}]{GKESB16}%
  \BibitemOpen
  \bibfield  {author} {\bibinfo {author} {\bibfnamefont {K.}~\bibnamefont
  {G\"{u}nthner}}, \bibinfo {author} {\bibfnamefont {I.}~\bibnamefont {Khan}},
  \bibinfo {author} {\bibfnamefont {D.}~\bibnamefont {Elser}}, \bibinfo
  {author} {\bibfnamefont {B.}~\bibnamefont {Stiller}}, \bibinfo {author}
  {\bibfnamefont {O.}~\bibnamefont {Bayraktar}}, \bibinfo {author}
  {\bibfnamefont {C.~R.}\ \bibnamefont {M\"{u}ller}}, \bibinfo {author}
  {\bibfnamefont {K.}~\bibnamefont {Sauke}}, \bibinfo {author} {\bibfnamefont
  {D.}~\bibnamefont {Tr\"{o}ndle}}, \bibinfo {author} {\bibfnamefont
  {F.}~\bibnamefont {Heine}}, \bibinfo {author} {\bibfnamefont
  {S.}~\bibnamefont {Seel}}, \bibinfo {author} {\bibfnamefont {P.}~\bibnamefont
  {Greulich}}, \bibinfo {author} {\bibfnamefont {H.}~\bibnamefont {Zech}},
  \bibinfo {author} {\bibfnamefont {B.}~\bibnamefont {G\"{u}tlich}}, \bibinfo
  {author} {\bibfnamefont {S.}~\bibnamefont {Philipp-May}}, \bibinfo {author}
  {\bibfnamefont {C.}~\bibnamefont {Marquardt}}, \ and\ \bibinfo {author}
  {\bibfnamefont {G.}~\bibnamefont {Leuchs}},\ }\href {\doibase
  10.1364/OPTICA.4.000611} {\bibfield  {journal} {\bibinfo  {journal} {Optica}\
  }\textbf {\bibinfo {volume} {4}},\ \bibinfo {pages} {611} (\bibinfo {year}
  {2017})}\BibitemShut {NoStop}%
\bibitem [{\citenamefont {Vedovato}\ \emph {et~al.}(2017)\citenamefont
  {Vedovato}, \citenamefont {Agnesi}, \citenamefont {Schiavon}, \citenamefont
  {Dequal}, \citenamefont {Calderaro}, \citenamefont {Tomasin}, \citenamefont
  {Giacomo~Marangon}, \citenamefont {Stanco}, \citenamefont {Luceri},
  \citenamefont {Bianco}, \citenamefont {Vallone},\ and\ \citenamefont
  {Villoresi}}]{Vedovato2017}%
  \BibitemOpen
  \bibfield  {author} {\bibinfo {author} {\bibfnamefont {F.}~\bibnamefont
  {Vedovato}}, \bibinfo {author} {\bibfnamefont {C.}~\bibnamefont {Agnesi}},
  \bibinfo {author} {\bibfnamefont {M.}~\bibnamefont {Schiavon}}, \bibinfo
  {author} {\bibfnamefont {D.}~\bibnamefont {Dequal}}, \bibinfo {author}
  {\bibfnamefont {L.}~\bibnamefont {Calderaro}}, \bibinfo {author}
  {\bibfnamefont {M.}~\bibnamefont {Tomasin}}, \bibinfo {author} {\bibfnamefont
  {D.}~\bibnamefont {Giacomo~Marangon}}, \bibinfo {author} {\bibfnamefont
  {A.}~\bibnamefont {Stanco}}, \bibinfo {author} {\bibfnamefont
  {V.}~\bibnamefont {Luceri}}, \bibinfo {author} {\bibfnamefont
  {G.}~\bibnamefont {Bianco}}, \bibinfo {author} {\bibfnamefont
  {G.}~\bibnamefont {Vallone}}, \ and\ \bibinfo {author} {\bibfnamefont
  {P.}~\bibnamefont {Villoresi}},\ }\href {\doibase
  https://doi.org/10.1126/sciadv.1701180} {\bibfield  {journal} {\bibinfo
  {journal} {Sci. Adv.}\ }\textbf {\bibinfo {volume} {3}} (\bibinfo {year}
  {2017}),\ https://doi.org/10.1126/sciadv.1701180}\BibitemShut {NoStop}%
\bibitem [{\citenamefont {Rideout}\ \emph {et~al.}(2012)\citenamefont
  {Rideout}, \citenamefont {Jennewein}, \citenamefont {Amelino-Camelia},
  \citenamefont {Demarie}, \citenamefont {Higgins}, \citenamefont {Kempf},
  \citenamefont {Kent}, \citenamefont {Laflamme}, \citenamefont {Ma},
  \citenamefont {Mann}, \citenamefont {Martin-Martinez}, \citenamefont
  {Menicucci}, \citenamefont {Moffat}, \citenamefont {Simon}, \citenamefont
  {Sorkin}, \citenamefont {Smolin},\ and\ \citenamefont {Terno}}]{Rideout2012}%
  \BibitemOpen
  \bibfield  {author} {\bibinfo {author} {\bibfnamefont {D.}~\bibnamefont
  {Rideout}}, \bibinfo {author} {\bibfnamefont {T.}~\bibnamefont {Jennewein}},
  \bibinfo {author} {\bibfnamefont {G.}~\bibnamefont {Amelino-Camelia}},
  \bibinfo {author} {\bibfnamefont {T.~F.}\ \bibnamefont {Demarie}}, \bibinfo
  {author} {\bibfnamefont {B.~L.}\ \bibnamefont {Higgins}}, \bibinfo {author}
  {\bibfnamefont {A.}~\bibnamefont {Kempf}}, \bibinfo {author} {\bibfnamefont
  {A.}~\bibnamefont {Kent}}, \bibinfo {author} {\bibfnamefont {R.}~\bibnamefont
  {Laflamme}}, \bibinfo {author} {\bibfnamefont {X.}~\bibnamefont {Ma}},
  \bibinfo {author} {\bibfnamefont {R.~B.}\ \bibnamefont {Mann}}, \bibinfo
  {author} {\bibfnamefont {E.}~\bibnamefont {Martin-Martinez}}, \bibinfo
  {author} {\bibfnamefont {N.~C.}\ \bibnamefont {Menicucci}}, \bibinfo {author}
  {\bibfnamefont {J.}~\bibnamefont {Moffat}}, \bibinfo {author} {\bibfnamefont
  {C.}~\bibnamefont {Simon}}, \bibinfo {author} {\bibfnamefont
  {R.}~\bibnamefont {Sorkin}}, \bibinfo {author} {\bibfnamefont
  {L.}~\bibnamefont {Smolin}}, \ and\ \bibinfo {author} {\bibfnamefont {D.~R.}\
  \bibnamefont {Terno}},\ }\href
  {http://stacks.iop.org/0264-9381/29/i=22/a=224011} {\bibfield  {journal}
  {\bibinfo  {journal} {Classical and Quantum Gravity}\ }\textbf {\bibinfo
  {volume} {29}},\ \bibinfo {pages} {224011} (\bibinfo {year}
  {2012})}\BibitemShut {NoStop}%
\bibitem [{\citenamefont {Weedbrook}\ \emph {et~al.}(2012)\citenamefont
  {Weedbrook}, \citenamefont {Pirandola}, \citenamefont {Garc\'{\i}a-Patr\'on},
  \citenamefont {Cerf}, \citenamefont {Ralph}, \citenamefont {Shapiro},\ and\
  \citenamefont {Lloyd}}]{WPGCRSL12}%
  \BibitemOpen
  \bibfield  {author} {\bibinfo {author} {\bibfnamefont {C.}~\bibnamefont
  {Weedbrook}}, \bibinfo {author} {\bibfnamefont {S.}~\bibnamefont
  {Pirandola}}, \bibinfo {author} {\bibfnamefont {R.}~\bibnamefont
  {Garc\'{\i}a-Patr\'on}}, \bibinfo {author} {\bibfnamefont {N.~J.}\
  \bibnamefont {Cerf}}, \bibinfo {author} {\bibfnamefont {T.~C.}\ \bibnamefont
  {Ralph}}, \bibinfo {author} {\bibfnamefont {J.~H.}\ \bibnamefont {Shapiro}},
  \ and\ \bibinfo {author} {\bibfnamefont {S.}~\bibnamefont {Lloyd}},\ }\href
  {\doibase 10.1103/RevModPhys.84.621} {\bibfield  {journal} {\bibinfo
  {journal} {Rev. Mod. Phys.}\ }\textbf {\bibinfo {volume} {84}},\ \bibinfo
  {pages} {621} (\bibinfo {year} {2012})}\BibitemShut {NoStop}%
\bibitem [{\citenamefont {Andersen}\ \emph {et~al.}(2016)\citenamefont
  {Andersen}, \citenamefont {Gehring}, \citenamefont {Marquardt},\ and\
  \citenamefont {Leuchs}}]{AGML16}%
  \BibitemOpen
  \bibfield  {author} {\bibinfo {author} {\bibfnamefont {U.~L.}\ \bibnamefont
  {Andersen}}, \bibinfo {author} {\bibfnamefont {T.}~\bibnamefont {Gehring}},
  \bibinfo {author} {\bibfnamefont {C.}~\bibnamefont {Marquardt}}, \ and\
  \bibinfo {author} {\bibfnamefont {G.}~\bibnamefont {Leuchs}},\ }\href
  {http://stacks.iop.org/1402-4896/91/i=5/a=053001} {\bibfield  {journal}
  {\bibinfo  {journal} {Physica Scripta}\ }\textbf {\bibinfo {volume} {91}},\
  \bibinfo {pages} {053001} (\bibinfo {year} {2016})}\BibitemShut {NoStop}%
\bibitem [{\citenamefont {Yuen}\ and\ \citenamefont {Chan}(1983)}]{YC83}%
  \BibitemOpen
  \bibfield  {author} {\bibinfo {author} {\bibfnamefont {H.~P.}\ \bibnamefont
  {Yuen}}\ and\ \bibinfo {author} {\bibfnamefont {V.~W.~S.}\ \bibnamefont
  {Chan}},\ }\href {\doibase 10.1364/OL.8.000177} {\bibfield  {journal}
  {\bibinfo  {journal} {Optics Letters}\ }\textbf {\bibinfo {volume} {8}},\
  \bibinfo {pages} {177} (\bibinfo {year} {1983})}\BibitemShut {NoStop}%
\bibitem [{\citenamefont {Aspelmeyer}\ \emph {et~al.}(2014)\citenamefont
  {Aspelmeyer}, \citenamefont {Kippenberg},\ and\ \citenamefont
  {Marquardt}}]{Marquardt2014}%
  \BibitemOpen
  \bibfield  {author} {\bibinfo {author} {\bibfnamefont {M.}~\bibnamefont
  {Aspelmeyer}}, \bibinfo {author} {\bibfnamefont {T.~J.}\ \bibnamefont
  {Kippenberg}}, \ and\ \bibinfo {author} {\bibfnamefont {F.}~\bibnamefont
  {Marquardt}},\ }\href {\doibase 10.1103/RevModPhys.86.1391} {\bibfield
  {journal} {\bibinfo  {journal} {Rev. Mod. Phys.}\ }\textbf {\bibinfo {volume}
  {86}},\ \bibinfo {pages} {1391} (\bibinfo {year} {2014})}\BibitemShut
  {NoStop}%
\bibitem [{\citenamefont {Houck}\ \emph {et~al.}(2007)\citenamefont {Houck},
  \citenamefont {Schuster}, \citenamefont {M.}, \citenamefont {Schreier},
  \citenamefont {Johnson}, \citenamefont {M.}, \citenamefont {Frunzio},
  \citenamefont {Majer}, \citenamefont {Devoret}, \citenamefont {Girvin},\ and\
  \citenamefont {J.}}]{Schoelkopf2007}%
  \BibitemOpen
  \bibfield  {author} {\bibinfo {author} {\bibfnamefont {A.~A.}\ \bibnamefont
  {Houck}}, \bibinfo {author} {\bibfnamefont {D.~I.}\ \bibnamefont {Schuster}},
  \bibinfo {author} {\bibfnamefont {G.~J.}\ \bibnamefont {M.}}, \bibinfo
  {author} {\bibfnamefont {J.~A.}\ \bibnamefont {Schreier}}, \bibinfo {author}
  {\bibfnamefont {B.~R.}\ \bibnamefont {Johnson}}, \bibinfo {author}
  {\bibfnamefont {C.~J.}\ \bibnamefont {M.}}, \bibinfo {author} {\bibfnamefont
  {L.}~\bibnamefont {Frunzio}}, \bibinfo {author} {\bibfnamefont
  {J.}~\bibnamefont {Majer}}, \bibinfo {author} {\bibfnamefont {M.~H.}\
  \bibnamefont {Devoret}}, \bibinfo {author} {\bibfnamefont {S.~M.}\
  \bibnamefont {Girvin}}, \ and\ \bibinfo {author} {\bibfnamefont {S.~R.}\
  \bibnamefont {J.}},\ }\href {\doibase doi:10.1038/nature06126} {\bibfield
  {journal} {\bibinfo  {journal} {Nature}\ }\textbf {\bibinfo {volume} {449}},\
  \bibinfo {pages} {328} (\bibinfo {year} {2007})}\BibitemShut {NoStop}%
\bibitem [{\citenamefont {Filipp}\ \emph {et~al.}(2009)\citenamefont {Filipp},
  \citenamefont {Maurer}, \citenamefont {Leek}, \citenamefont {Baur},
  \citenamefont {Bianchetti}, \citenamefont {Fink}, \citenamefont {G\"oppl},
  \citenamefont {Steffen}, \citenamefont {Gambetta}, \citenamefont {Blais},\
  and\ \citenamefont {Wallraff}}]{Wallraff2009}%
  \BibitemOpen
  \bibfield  {author} {\bibinfo {author} {\bibfnamefont {S.}~\bibnamefont
  {Filipp}}, \bibinfo {author} {\bibfnamefont {P.}~\bibnamefont {Maurer}},
  \bibinfo {author} {\bibfnamefont {P.~J.}\ \bibnamefont {Leek}}, \bibinfo
  {author} {\bibfnamefont {M.}~\bibnamefont {Baur}}, \bibinfo {author}
  {\bibfnamefont {R.}~\bibnamefont {Bianchetti}}, \bibinfo {author}
  {\bibfnamefont {J.~M.}\ \bibnamefont {Fink}}, \bibinfo {author}
  {\bibfnamefont {M.}~\bibnamefont {G\"oppl}}, \bibinfo {author} {\bibfnamefont
  {L.}~\bibnamefont {Steffen}}, \bibinfo {author} {\bibfnamefont {J.~M.}\
  \bibnamefont {Gambetta}}, \bibinfo {author} {\bibfnamefont {A.}~\bibnamefont
  {Blais}}, \ and\ \bibinfo {author} {\bibfnamefont {A.}~\bibnamefont
  {Wallraff}},\ }\href {\doibase 10.1103/PhysRevLett.102.200402} {\bibfield
  {journal} {\bibinfo  {journal} {Phys. Rev. Lett.}\ }\textbf {\bibinfo
  {volume} {102}},\ \bibinfo {pages} {200402} (\bibinfo {year}
  {2009})}\BibitemShut {NoStop}%
\bibitem [{\citenamefont {Kuzmich}\ and\ \citenamefont {Polzik}(2003)}]{KP03}%
  \BibitemOpen
  \bibfield  {author} {\bibinfo {author} {\bibfnamefont {A.}~\bibnamefont
  {Kuzmich}}\ and\ \bibinfo {author} {\bibfnamefont {E.~S.}\ \bibnamefont
  {Polzik}},\ }in\ \href@noop {} {\emph {\bibinfo {booktitle} {Quantum
  Information with Continuous Variables}}}\ (\bibinfo  {publisher} {Springer},\
  \bibinfo {year} {2003})\ pp.\ \bibinfo {pages} {231--265}\BibitemShut
  {NoStop}%
\bibitem [{\citenamefont {Cerf}\ \emph {et~al.}(2007)\citenamefont {Cerf},
  \citenamefont {Leuchs},\ and\ \citenamefont {Polzik}}]{CLP07}%
  \BibitemOpen
  \bibfield  {author} {\bibinfo {author} {\bibfnamefont {N.~J.}\ \bibnamefont
  {Cerf}}, \bibinfo {author} {\bibfnamefont {G.}~\bibnamefont {Leuchs}}, \ and\
  \bibinfo {author} {\bibfnamefont {E.~S.}\ \bibnamefont {Polzik}},\
  }\href@noop {} {\emph {\bibinfo {title} {Quantum information with continuous
  variables of atoms and light}}}\ (\bibinfo  {publisher} {Imperial College
  Press},\ \bibinfo {year} {2007})\BibitemShut {NoStop}%
\bibitem [{\citenamefont {Namiki}\ \emph {et~al.}(2011)\citenamefont {Namiki},
  \citenamefont {Tanaka}, \citenamefont {Takano},\ and\ \citenamefont
  {Takahashi}}]{NTTT11}%
  \BibitemOpen
  \bibfield  {author} {\bibinfo {author} {\bibfnamefont {R.}~\bibnamefont
  {Namiki}}, \bibinfo {author} {\bibfnamefont {S.~I.~R.}\ \bibnamefont
  {Tanaka}}, \bibinfo {author} {\bibfnamefont {T.}~\bibnamefont {Takano}}, \
  and\ \bibinfo {author} {\bibfnamefont {Y.}~\bibnamefont {Takahashi}},\ }\href
  {\doibase 10.1007/s00340-011-4717-7} {\bibfield  {journal} {\bibinfo
  {journal} {Applied Physics B}\ }\textbf {\bibinfo {volume} {105}},\ \bibinfo
  {pages} {197} (\bibinfo {year} {2011})}\BibitemShut {NoStop}%
\bibitem [{\citenamefont {M\"uller}\ \emph
  {et~al.}(2016{\natexlab{a}})\citenamefont {M\"uller}, \citenamefont {Madsen},
  \citenamefont {Klimov}, \citenamefont {S\'anchez-Soto}, \citenamefont
  {Leuchs}, \citenamefont {Marquardt},\ and\ \citenamefont
  {Andersen}}]{MMKS16}%
  \BibitemOpen
  \bibfield  {author} {\bibinfo {author} {\bibfnamefont {C.~R.}\ \bibnamefont
  {M\"uller}}, \bibinfo {author} {\bibfnamefont {L.~S.}\ \bibnamefont
  {Madsen}}, \bibinfo {author} {\bibfnamefont {A.~B.}\ \bibnamefont {Klimov}},
  \bibinfo {author} {\bibfnamefont {L.~L.}\ \bibnamefont {S\'anchez-Soto}},
  \bibinfo {author} {\bibfnamefont {G.}~\bibnamefont {Leuchs}}, \bibinfo
  {author} {\bibfnamefont {C.}~\bibnamefont {Marquardt}}, \ and\ \bibinfo
  {author} {\bibfnamefont {U.~L.}\ \bibnamefont {Andersen}},\ }\href {\doibase
  10.1103/PhysRevA.93.033816} {\bibfield  {journal} {\bibinfo  {journal} {Phys.
  Rev. A}\ }\textbf {\bibinfo {volume} {93}},\ \bibinfo {pages} {033816}
  (\bibinfo {year} {2016}{\natexlab{a}})}\BibitemShut {NoStop}%
\bibitem [{\citenamefont {Est{\'e}ve}\ \emph {et~al.}(2008)\citenamefont
  {Est{\'e}ve}, \citenamefont {Gross}, \citenamefont {Weller}, \citenamefont
  {Giovanazzi},\ and\ \citenamefont {K.}}]{Oberthaler2008}%
  \BibitemOpen
  \bibfield  {author} {\bibinfo {author} {\bibfnamefont {J.}~\bibnamefont
  {Est{\'e}ve}}, \bibinfo {author} {\bibfnamefont {C.}~\bibnamefont {Gross}},
  \bibinfo {author} {\bibfnamefont {A.}~\bibnamefont {Weller}}, \bibinfo
  {author} {\bibfnamefont {S.}~\bibnamefont {Giovanazzi}}, \ and\ \bibinfo
  {author} {\bibfnamefont {O.~M.}\ \bibnamefont {K.}},\ }\href {\doibase
  doi:10.1038/nature07332} {\bibfield  {journal} {\bibinfo  {journal} {Nature}\
  }\textbf {\bibinfo {volume} {455}},\ \bibinfo {pages} {1216} (\bibinfo {year}
  {2008})}\BibitemShut {NoStop}%
\bibitem [{\citenamefont {Braunstein}\ and\ \citenamefont {van
  Loock}(2005)}]{BvL05}%
  \BibitemOpen
  \bibfield  {author} {\bibinfo {author} {\bibfnamefont {S.~L.}\ \bibnamefont
  {Braunstein}}\ and\ \bibinfo {author} {\bibfnamefont {P.}~\bibnamefont {van
  Loock}},\ }\href {\doibase 10.1103/RevModPhys.77.513} {\bibfield  {journal}
  {\bibinfo  {journal} {Rev. Mod. Phys.}\ }\textbf {\bibinfo {volume} {77}},\
  \bibinfo {pages} {513} (\bibinfo {year} {2005})}\BibitemShut {NoStop}%
\bibitem [{\citenamefont {Smithey}\ \emph {et~al.}(1993)\citenamefont
  {Smithey}, \citenamefont {Beck}, \citenamefont {Raymer},\ and\ \citenamefont
  {Faridani}}]{SBRF93}%
  \BibitemOpen
  \bibfield  {author} {\bibinfo {author} {\bibfnamefont {D.~T.}\ \bibnamefont
  {Smithey}}, \bibinfo {author} {\bibfnamefont {M.}~\bibnamefont {Beck}},
  \bibinfo {author} {\bibfnamefont {M.~G.}\ \bibnamefont {Raymer}}, \ and\
  \bibinfo {author} {\bibfnamefont {A.}~\bibnamefont {Faridani}},\ }\href
  {\doibase https://doi.org/10.1103/PhysRevLett.70.1244} {\bibfield  {journal}
  {\bibinfo  {journal} {Phys. Rev. Lett.}\ }\textbf {\bibinfo {volume} {70}},\
  \bibinfo {pages} {1244} (\bibinfo {year} {1993})}\BibitemShut {NoStop}%
\bibitem [{\citenamefont {Leonhardt}(1997)}]{Leo97}%
  \BibitemOpen
  \bibfield  {author} {\bibinfo {author} {\bibfnamefont {U.}~\bibnamefont
  {Leonhardt}},\ }\href@noop {} {\emph {\bibinfo {title} {Measuring the quantum
  state of light}}},\ Vol.~\bibinfo {volume} {22}\ (\bibinfo  {publisher}
  {Cambridge university press},\ \bibinfo {year} {1997})\BibitemShut {NoStop}%
\bibitem [{\citenamefont {Lvovsky}\ and\ \citenamefont {Raymer}(2009)}]{LR09}%
  \BibitemOpen
  \bibfield  {author} {\bibinfo {author} {\bibfnamefont {A.~I.}\ \bibnamefont
  {Lvovsky}}\ and\ \bibinfo {author} {\bibfnamefont {M.~G.}\ \bibnamefont
  {Raymer}},\ }\href {\doibase 10.1103/RevModPhys.81.299} {\bibfield  {journal}
  {\bibinfo  {journal} {Rev. Mod. Phys.}\ }\textbf {\bibinfo {volume} {81}},\
  \bibinfo {pages} {299} (\bibinfo {year} {2009})}\BibitemShut {NoStop}%
\bibitem [{\citenamefont {M\"uller}\ \emph
  {et~al.}(2016{\natexlab{b}})\citenamefont {M\"uller}, \citenamefont
  {Peuntinger}, \citenamefont {Dirmeier}, \citenamefont {Khan}, \citenamefont
  {Vogl}, \citenamefont {Marquardt}, \citenamefont {Leuchs}, \citenamefont
  {S\'anchez-Soto}, \citenamefont {Teo}, \citenamefont {Hradil},\ and\
  \citenamefont {\ifmmode \check{R}\else \v{R}\fi{}eh\'a\ifmmode~\check{c}\else
  \v{c}\fi{}ek}}]{MPDKV16}%
  \BibitemOpen
  \bibfield  {author} {\bibinfo {author} {\bibfnamefont {C.~R.}\ \bibnamefont
  {M\"uller}}, \bibinfo {author} {\bibfnamefont {C.}~\bibnamefont
  {Peuntinger}}, \bibinfo {author} {\bibfnamefont {T.}~\bibnamefont
  {Dirmeier}}, \bibinfo {author} {\bibfnamefont {I.}~\bibnamefont {Khan}},
  \bibinfo {author} {\bibfnamefont {U.}~\bibnamefont {Vogl}}, \bibinfo {author}
  {\bibfnamefont {C.}~\bibnamefont {Marquardt}}, \bibinfo {author}
  {\bibfnamefont {G.}~\bibnamefont {Leuchs}}, \bibinfo {author} {\bibfnamefont
  {L.~L.}\ \bibnamefont {S\'anchez-Soto}}, \bibinfo {author} {\bibfnamefont
  {Y.~S.}\ \bibnamefont {Teo}}, \bibinfo {author} {\bibfnamefont
  {Z.}~\bibnamefont {Hradil}}, \ and\ \bibinfo {author} {\bibfnamefont
  {J.}~\bibnamefont {\ifmmode \check{R}\else
  \v{R}\fi{}eh\'a\ifmmode~\check{c}\else \v{c}\fi{}ek}},\ }\href {\doibase
  10.1103/PhysRevLett.117.070801} {\bibfield  {journal} {\bibinfo  {journal}
  {Phys. Rev. Lett.}\ }\textbf {\bibinfo {volume} {117}},\ \bibinfo {pages}
  {070801} (\bibinfo {year} {2016}{\natexlab{b}})}\BibitemShut {NoStop}%
\bibitem [{\citenamefont {Zhao}\ \emph {et~al.}(2009)\citenamefont {Zhao},
  \citenamefont {Heid}, \citenamefont {Rigas},\ and\ \citenamefont
  {L\"utkenhaus}}]{ZHRL09}%
  \BibitemOpen
  \bibfield  {author} {\bibinfo {author} {\bibfnamefont {Y.-B.}\ \bibnamefont
  {Zhao}}, \bibinfo {author} {\bibfnamefont {M.}~\bibnamefont {Heid}}, \bibinfo
  {author} {\bibfnamefont {J.}~\bibnamefont {Rigas}}, \ and\ \bibinfo {author}
  {\bibfnamefont {N.}~\bibnamefont {L\"utkenhaus}},\ }\href {\doibase
  10.1103/PhysRevA.79.012307} {\bibfield  {journal} {\bibinfo  {journal} {Phys.
  Rev. A}\ }\textbf {\bibinfo {volume} {79}},\ \bibinfo {pages} {012307}
  (\bibinfo {year} {2009})}\BibitemShut {NoStop}%
\bibitem [{\citenamefont {Gilchrist}\ \emph {et~al.}(1999)\citenamefont
  {Gilchrist}, \citenamefont {Deuar},\ and\ \citenamefont {Reid}}]{GDR99}%
  \BibitemOpen
  \bibfield  {author} {\bibinfo {author} {\bibfnamefont {A.}~\bibnamefont
  {Gilchrist}}, \bibinfo {author} {\bibfnamefont {P.}~\bibnamefont {Deuar}}, \
  and\ \bibinfo {author} {\bibfnamefont {M.~D.}\ \bibnamefont {Reid}},\ }\href
  {\doibase 10.1103/PhysRevA.60.4259} {\bibfield  {journal} {\bibinfo
  {journal} {Phys. Rev. A}\ }\textbf {\bibinfo {volume} {60}},\ \bibinfo
  {pages} {4259} (\bibinfo {year} {1999})}\BibitemShut {NoStop}%
\bibitem [{\citenamefont {Munro}(1999)}]{MCVBell99}%
  \BibitemOpen
  \bibfield  {author} {\bibinfo {author} {\bibfnamefont {W.~J.}\ \bibnamefont
  {Munro}},\ }\href {\doibase 10.1103/PhysRevA.59.4197} {\bibfield  {journal}
  {\bibinfo  {journal} {Phys. Rev. A}\ }\textbf {\bibinfo {volume} {59}},\
  \bibinfo {pages} {4197} (\bibinfo {year} {1999})}\BibitemShut {NoStop}%
\bibitem [{\citenamefont {Auberson}\ \emph {et~al.}(2002)\citenamefont
  {Auberson}, \citenamefont {Mahoux}, \citenamefont {Roy},\ and\ \citenamefont
  {Singh}}]{AUBERSON2002327}%
  \BibitemOpen
  \bibfield  {author} {\bibinfo {author} {\bibfnamefont {G.}~\bibnamefont
  {Auberson}}, \bibinfo {author} {\bibfnamefont {G.}~\bibnamefont {Mahoux}},
  \bibinfo {author} {\bibfnamefont {S.}~\bibnamefont {Roy}}, \ and\ \bibinfo
  {author} {\bibfnamefont {V.}~\bibnamefont {Singh}},\ }\href {\doibase
  http://dx.doi.org/10.1016/S0375-9601(02)00827-7} {\bibfield  {journal}
  {\bibinfo  {journal} {Physics Letters A}\ }\textbf {\bibinfo {volume}
  {300}},\ \bibinfo {pages} {327 } (\bibinfo {year} {2002})}\BibitemShut
  {NoStop}%
\bibitem [{\citenamefont {Banaszek}\ and\ \citenamefont
  {W\'odkiewicz}(1999)}]{BW99}%
  \BibitemOpen
  \bibfield  {author} {\bibinfo {author} {\bibfnamefont {K.}~\bibnamefont
  {Banaszek}}\ and\ \bibinfo {author} {\bibfnamefont {K.}~\bibnamefont
  {W\'odkiewicz}},\ }\href {\doibase 10.1103/PhysRevLett.82.2009} {\bibfield
  {journal} {\bibinfo  {journal} {Phys. Rev. Lett.}\ }\textbf {\bibinfo
  {volume} {82}},\ \bibinfo {pages} {2009} (\bibinfo {year}
  {1999})}\BibitemShut {NoStop}%
\bibitem [{\citenamefont {Seshadreesan}\ \emph {et~al.}(2016)\citenamefont
  {Seshadreesan}, \citenamefont {Wildfeuer}, \citenamefont {Kim}, \citenamefont
  {Lee},\ and\ \citenamefont {Dowling}}]{SWKLD16}%
  \BibitemOpen
  \bibfield  {author} {\bibinfo {author} {\bibfnamefont {K.~P.}\ \bibnamefont
  {Seshadreesan}}, \bibinfo {author} {\bibfnamefont {C.~F.}\ \bibnamefont
  {Wildfeuer}}, \bibinfo {author} {\bibfnamefont {M.~B.}\ \bibnamefont {Kim}},
  \bibinfo {author} {\bibfnamefont {H.}~\bibnamefont {Lee}}, \ and\ \bibinfo
  {author} {\bibfnamefont {J.~P.}\ \bibnamefont {Dowling}},\ }\href {\doibase
  10.1007/s11128-015-1082-1} {\bibfield  {journal} {\bibinfo  {journal}
  {Quantum Information Processing}\ }\textbf {\bibinfo {volume} {15}},\
  \bibinfo {pages} {1025} (\bibinfo {year} {2016})}\BibitemShut {NoStop}%
\bibitem [{\citenamefont {Schaefer}\ \emph {et~al.}(2016)\citenamefont
  {Schaefer}, \citenamefont {Gregory},\ and\ \citenamefont
  {Rosenkranz}}]{SGR16}%
  \BibitemOpen
  \bibfield  {author} {\bibinfo {author} {\bibfnamefont {S.}~\bibnamefont
  {Schaefer}}, \bibinfo {author} {\bibfnamefont {M.}~\bibnamefont {Gregory}}, \
  and\ \bibinfo {author} {\bibfnamefont {W.}~\bibnamefont {Rosenkranz}},\
  }\href {https://doi.org/10.1117/1.OE.55.11.111614} {\bibfield  {journal}
  {\bibinfo  {journal} {Opt. Eng.}\ }\textbf {\bibinfo {volume} {55(11)}},\
  \bibinfo {pages} {111614} (\bibinfo {year} {2016})}\BibitemShut {NoStop}%
\bibitem [{\citenamefont {Heine}\ \emph {et~al.}(2015)\citenamefont {Heine},
  \citenamefont {Mühlnikel}, \citenamefont {Zech}, \citenamefont {Tröndle},
  \citenamefont {Seel}, \citenamefont {Motzigemba}, \citenamefont {Meyer},
  \citenamefont {Philipp-May},\ and\ \citenamefont {Benzi}}]{HMZ15}%
  \BibitemOpen
  \bibfield  {author} {\bibinfo {author} {\bibfnamefont {F.}~\bibnamefont
  {Heine}}, \bibinfo {author} {\bibfnamefont {G.}~\bibnamefont {Mühlnikel}},
  \bibinfo {author} {\bibfnamefont {H.}~\bibnamefont {Zech}}, \bibinfo {author}
  {\bibfnamefont {D.}~\bibnamefont {Tröndle}}, \bibinfo {author} {\bibfnamefont
  {S.}~\bibnamefont {Seel}}, \bibinfo {author} {\bibfnamefont {M.}~\bibnamefont
  {Motzigemba}}, \bibinfo {author} {\bibfnamefont {R.}~\bibnamefont {Meyer}},
  \bibinfo {author} {\bibfnamefont {S.}~\bibnamefont {Philipp-May}}, \ and\
  \bibinfo {author} {\bibfnamefont {E.}~\bibnamefont {Benzi}},\ }\href
  {\doibase 10.1117/12.2083117} {\bibfield  {journal} {\bibinfo  {journal}
  {Proc. SPIE}\ }\textbf {\bibinfo {volume} {93540G}} (\bibinfo {year}
  {2015}),\ 10.1117/12.2083117}\BibitemShut {NoStop}%
\bibitem [{\citenamefont {Dolinar}(1973)}]{Do73}%
  \BibitemOpen
  \bibfield  {author} {\bibinfo {author} {\bibfnamefont {S.~J.}\ \bibnamefont
  {Dolinar}},\ }\href@noop {} {\emph {\bibinfo {title} {An optimum receiver for
  the binary coherent state quantum channel.}}},\ \bibinfo {type} {Quarterly
  Progress Report}\ \bibinfo {number} {111: 115-120.}\ (\bibinfo  {institution}
  {MIT Research Laboratory of Electronics},\ \bibinfo {year}
  {1973})\BibitemShut {NoStop}%
\bibitem [{\citenamefont {Wittmann}\ \emph {et~al.}(2008)\citenamefont
  {Wittmann}, \citenamefont {Takeoka}, \citenamefont {Cassemiro}, \citenamefont
  {Sasaki}, \citenamefont {Leuchs},\ and\ \citenamefont {Andersen}}]{WTCSLA08}%
  \BibitemOpen
  \bibfield  {author} {\bibinfo {author} {\bibfnamefont {C.}~\bibnamefont
  {Wittmann}}, \bibinfo {author} {\bibfnamefont {M.}~\bibnamefont {Takeoka}},
  \bibinfo {author} {\bibfnamefont {K.~N.}\ \bibnamefont {Cassemiro}}, \bibinfo
  {author} {\bibfnamefont {M.}~\bibnamefont {Sasaki}}, \bibinfo {author}
  {\bibfnamefont {G.}~\bibnamefont {Leuchs}}, \ and\ \bibinfo {author}
  {\bibfnamefont {U.~L.}\ \bibnamefont {Andersen}},\ }\href {\doibase
  10.1103/PhysRevLett.101.210501} {\bibfield  {journal} {\bibinfo  {journal}
  {Phys. Rev. Lett.}\ }\textbf {\bibinfo {volume} {101}},\ \bibinfo {pages}
  {210501} (\bibinfo {year} {2008})}\BibitemShut {NoStop}%
\bibitem [{\citenamefont {Takeoka}\ and\ \citenamefont {Sasaki}(2008)}]{TS08}%
  \BibitemOpen
  \bibfield  {author} {\bibinfo {author} {\bibfnamefont {M.}~\bibnamefont
  {Takeoka}}\ and\ \bibinfo {author} {\bibfnamefont {M.}~\bibnamefont
  {Sasaki}},\ }\href {\doibase 10.1103/PhysRevA.78.022320} {\bibfield
  {journal} {\bibinfo  {journal} {Phys. Rev. A}\ }\textbf {\bibinfo {volume}
  {78}},\ \bibinfo {pages} {022320} (\bibinfo {year} {2008})}\BibitemShut
  {NoStop}%
\bibitem [{\citenamefont {Wittmann}\ \emph {et~al.}(2010)\citenamefont
  {Wittmann}, \citenamefont {Andersen}, \citenamefont {Takeoka}, \citenamefont
  {Sych},\ and\ \citenamefont {Leuchs}}]{WATSL10}%
  \BibitemOpen
  \bibfield  {author} {\bibinfo {author} {\bibfnamefont {C.}~\bibnamefont
  {Wittmann}}, \bibinfo {author} {\bibfnamefont {U.~L.}\ \bibnamefont
  {Andersen}}, \bibinfo {author} {\bibfnamefont {M.}~\bibnamefont {Takeoka}},
  \bibinfo {author} {\bibfnamefont {D.}~\bibnamefont {Sych}}, \ and\ \bibinfo
  {author} {\bibfnamefont {G.}~\bibnamefont {Leuchs}},\ }\href {\doibase
  10.1103/PhysRevLett.104.100505} {\bibfield  {journal} {\bibinfo  {journal}
  {Phys. Rev. Lett.}\ }\textbf {\bibinfo {volume} {104}},\ \bibinfo {pages}
  {100505} (\bibinfo {year} {2010})}\BibitemShut {NoStop}%
\bibitem [{\citenamefont {Wilde}\ \emph {et~al.}(2012)\citenamefont {Wilde},
  \citenamefont {Guha}, \citenamefont {Tan},\ and\ \citenamefont
  {Lloyd}}]{WGTL12}%
  \BibitemOpen
  \bibfield  {author} {\bibinfo {author} {\bibfnamefont {M.~M.}\ \bibnamefont
  {Wilde}}, \bibinfo {author} {\bibfnamefont {S.}~\bibnamefont {Guha}},
  \bibinfo {author} {\bibfnamefont {S.~H.}\ \bibnamefont {Tan}}, \ and\
  \bibinfo {author} {\bibfnamefont {S.}~\bibnamefont {Lloyd}},\ }in\ \href
  {\doibase 10.1109/ISIT.2012.6284251} {\emph {\bibinfo {booktitle}
  {Information Theory Proceedings (ISIT), 2012 IEEE International Symposium
  on}}}\ (\bibinfo {year} {2012})\ pp.\ \bibinfo {pages} {551--555}\BibitemShut
  {NoStop}%
\bibitem [{\citenamefont {Dowling}\ and\ \citenamefont
  {Seshadreesan}(2015)}]{DS15}%
  \BibitemOpen
  \bibfield  {author} {\bibinfo {author} {\bibfnamefont {J.~P.}\ \bibnamefont
  {Dowling}}\ and\ \bibinfo {author} {\bibfnamefont {K.~P.}\ \bibnamefont
  {Seshadreesan}},\ }\href {http://jlt.osa.org/abstract.cfm?URI=jlt-33-12-2359}
  {\bibfield  {journal} {\bibinfo  {journal} {J. Lightwave Technol.}\ }\textbf
  {\bibinfo {volume} {33}},\ \bibinfo {pages} {2359} (\bibinfo {year}
  {2015})}\BibitemShut {NoStop}%
\bibitem [{\citenamefont {Demkowicz-Dobrza\'nski}\ \emph
  {et~al.}(2015)\citenamefont {Demkowicz-Dobrza\'nski}, \citenamefont
  {Jarzyna},\ and\ \citenamefont {Ko{\l}ody\'nski}}]{DJK15}%
  \BibitemOpen
  \bibfield  {author} {\bibinfo {author} {\bibfnamefont {R.}~\bibnamefont
  {Demkowicz-Dobrza\'nski}}, \bibinfo {author} {\bibfnamefont {M.}~\bibnamefont
  {Jarzyna}}, \ and\ \bibinfo {author} {\bibfnamefont {J.}~\bibnamefont
  {Ko{\l}ody\'nski}},\ }\href {\doibase
  http://dx.doi.org/10.1016/bs.po.2015.02.003} {\bibfield  {journal} {\bibinfo
  {journal} {Progress in Optics}\ }\textbf {\bibinfo {volume} {60}},\ \bibinfo
  {pages} {345 } (\bibinfo {year} {2015})}\BibitemShut {NoStop}%
\bibitem [{\citenamefont {Seshadreesan}\ \emph {et~al.}(2013)\citenamefont
  {Seshadreesan}, \citenamefont {Kim}, \citenamefont {Dowling},\ and\
  \citenamefont {Lee}}]{SKDL13}%
  \BibitemOpen
  \bibfield  {author} {\bibinfo {author} {\bibfnamefont {K.~P.}\ \bibnamefont
  {Seshadreesan}}, \bibinfo {author} {\bibfnamefont {S.}~\bibnamefont {Kim}},
  \bibinfo {author} {\bibfnamefont {J.~P.}\ \bibnamefont {Dowling}}, \ and\
  \bibinfo {author} {\bibfnamefont {H.}~\bibnamefont {Lee}},\ }\href {\doibase
  10.1103/PhysRevA.87.043833} {\bibfield  {journal} {\bibinfo  {journal} {Phys.
  Rev. A}\ }\textbf {\bibinfo {volume} {87}},\ \bibinfo {pages} {043833}
  (\bibinfo {year} {2013})}\BibitemShut {NoStop}%
\bibitem [{\citenamefont {Seshadreesan}\ \emph {et~al.}(2011)\citenamefont
  {Seshadreesan}, \citenamefont {Anisimov}, \citenamefont {Lee},\ and\
  \citenamefont {Dowling}}]{SALD11}%
  \BibitemOpen
  \bibfield  {author} {\bibinfo {author} {\bibfnamefont {K.~P.}\ \bibnamefont
  {Seshadreesan}}, \bibinfo {author} {\bibfnamefont {P.~M.}\ \bibnamefont
  {Anisimov}}, \bibinfo {author} {\bibfnamefont {H.}~\bibnamefont {Lee}}, \
  and\ \bibinfo {author} {\bibfnamefont {J.~P.}\ \bibnamefont {Dowling}},\
  }\href {http://stacks.iop.org/1367-2630/13/i=8/a=083026} {\bibfield
  {journal} {\bibinfo  {journal} {New Journal of Physics}\ }\textbf {\bibinfo
  {volume} {13}},\ \bibinfo {pages} {083026} (\bibinfo {year}
  {2011})}\BibitemShut {NoStop}%
\bibitem [{\citenamefont {Anisimov}\ \emph {et~al.}(2010)\citenamefont
  {Anisimov}, \citenamefont {Raterman}, \citenamefont {Chiruvelli},
  \citenamefont {Plick}, \citenamefont {Huver}, \citenamefont {Lee},\ and\
  \citenamefont {Dowling}}]{ARCPH10}%
  \BibitemOpen
  \bibfield  {author} {\bibinfo {author} {\bibfnamefont {P.~M.}\ \bibnamefont
  {Anisimov}}, \bibinfo {author} {\bibfnamefont {G.~M.}\ \bibnamefont
  {Raterman}}, \bibinfo {author} {\bibfnamefont {A.}~\bibnamefont
  {Chiruvelli}}, \bibinfo {author} {\bibfnamefont {W.~N.}\ \bibnamefont
  {Plick}}, \bibinfo {author} {\bibfnamefont {S.~D.}\ \bibnamefont {Huver}},
  \bibinfo {author} {\bibfnamefont {H.}~\bibnamefont {Lee}}, \ and\ \bibinfo
  {author} {\bibfnamefont {J.~P.}\ \bibnamefont {Dowling}},\ }\href {\doibase
  10.1103/PhysRevLett.104.103602} {\bibfield  {journal} {\bibinfo  {journal}
  {Phys. Rev. Lett.}\ }\textbf {\bibinfo {volume} {104}},\ \bibinfo {pages}
  {103602} (\bibinfo {year} {2010})}\BibitemShut {NoStop}%
\bibitem [{\citenamefont {Motes}\ \emph {et~al.}(2015)\citenamefont {Motes},
  \citenamefont {Olson}, \citenamefont {Rabeaux}, \citenamefont {Dowling},
  \citenamefont {Olson},\ and\ \citenamefont {Rohde}}]{MORDOR15}%
  \BibitemOpen
  \bibfield  {author} {\bibinfo {author} {\bibfnamefont {K.~R.}\ \bibnamefont
  {Motes}}, \bibinfo {author} {\bibfnamefont {J.~P.}\ \bibnamefont {Olson}},
  \bibinfo {author} {\bibfnamefont {E.~J.}\ \bibnamefont {Rabeaux}}, \bibinfo
  {author} {\bibfnamefont {J.~P.}\ \bibnamefont {Dowling}}, \bibinfo {author}
  {\bibfnamefont {S.~J.}\ \bibnamefont {Olson}}, \ and\ \bibinfo {author}
  {\bibfnamefont {P.~P.}\ \bibnamefont {Rohde}},\ }\href {\doibase
  10.1103/PhysRevLett.114.170802} {\bibfield  {journal} {\bibinfo  {journal}
  {Phys. Rev. Lett.}\ }\textbf {\bibinfo {volume} {114}},\ \bibinfo {pages}
  {170802} (\bibinfo {year} {2015})}\BibitemShut {NoStop}%
\bibitem [{\citenamefont {Morin}\ \emph {et~al.}(2013)\citenamefont {Morin},
  \citenamefont {Bancal}, \citenamefont {Ho}, \citenamefont {Sekatski},
  \citenamefont {D'Auria}, \citenamefont {Gisin}, \citenamefont {Laurat},\ and\
  \citenamefont {Sangouard}}]{MBHSD13}%
  \BibitemOpen
  \bibfield  {author} {\bibinfo {author} {\bibfnamefont {O.}~\bibnamefont
  {Morin}}, \bibinfo {author} {\bibfnamefont {J.-D.}\ \bibnamefont {Bancal}},
  \bibinfo {author} {\bibfnamefont {M.}~\bibnamefont {Ho}}, \bibinfo {author}
  {\bibfnamefont {P.}~\bibnamefont {Sekatski}}, \bibinfo {author}
  {\bibfnamefont {V.}~\bibnamefont {D'Auria}}, \bibinfo {author} {\bibfnamefont
  {N.}~\bibnamefont {Gisin}}, \bibinfo {author} {\bibfnamefont
  {J.}~\bibnamefont {Laurat}}, \ and\ \bibinfo {author} {\bibfnamefont
  {N.}~\bibnamefont {Sangouard}},\ }\href {\doibase
  10.1103/PhysRevLett.110.130401} {\bibfield  {journal} {\bibinfo  {journal}
  {Phys. Rev. Lett.}\ }\textbf {\bibinfo {volume} {110}},\ \bibinfo {pages}
  {130401} (\bibinfo {year} {2013})}\BibitemShut {NoStop}%
\bibitem [{\citenamefont {Distante}\ \emph {et~al.}(2013)\citenamefont
  {Distante}, \citenamefont {Je\ifmmode~\check{z}\else \v{z}\fi{}ek},\ and\
  \citenamefont {Andersen}}]{DJA13}%
  \BibitemOpen
  \bibfield  {author} {\bibinfo {author} {\bibfnamefont {E.}~\bibnamefont
  {Distante}}, \bibinfo {author} {\bibfnamefont {M.}~\bibnamefont
  {Je\ifmmode~\check{z}\else \v{z}\fi{}ek}}, \ and\ \bibinfo {author}
  {\bibfnamefont {U.~L.}\ \bibnamefont {Andersen}},\ }\href {\doibase
  10.1103/PhysRevLett.111.033603} {\bibfield  {journal} {\bibinfo  {journal}
  {Phys. Rev. Lett.}\ }\textbf {\bibinfo {volume} {111}},\ \bibinfo {pages}
  {033603} (\bibinfo {year} {2013})}\BibitemShut {NoStop}%
\bibitem [{\citenamefont {Heersink}\ \emph {et~al.}(2005)\citenamefont
  {Heersink}, \citenamefont {Josse}, \citenamefont {Leuchs},\ and\
  \citenamefont {Andersen}}]{HJLA2005}%
  \BibitemOpen
  \bibfield  {author} {\bibinfo {author} {\bibfnamefont {J.}~\bibnamefont
  {Heersink}}, \bibinfo {author} {\bibfnamefont {V.}~\bibnamefont {Josse}},
  \bibinfo {author} {\bibfnamefont {G.}~\bibnamefont {Leuchs}}, \ and\ \bibinfo
  {author} {\bibfnamefont {U.~L.}\ \bibnamefont {Andersen}},\ }\href {\doibase
  https://doi.org/10.1364/OL.30.001192} {\bibfield  {journal} {\bibinfo
  {journal} {Opt. Lett.}\ }\textbf {\bibinfo {volume} {30}},\ \bibinfo {pages}
  {1192} (\bibinfo {year} {2005})}\BibitemShut {NoStop}%
\bibitem [{\citenamefont {Kitagawa}\ and\ \citenamefont
  {Yamamoto}(1986)}]{KY86}%
  \BibitemOpen
  \bibfield  {author} {\bibinfo {author} {\bibfnamefont {M.}~\bibnamefont
  {Kitagawa}}\ and\ \bibinfo {author} {\bibfnamefont {Y.}~\bibnamefont
  {Yamamoto}},\ }\href {\doibase 10.1103/PhysRevA.34.3974} {\bibfield
  {journal} {\bibinfo  {journal} {Phys. Rev. A}\ }\textbf {\bibinfo {volume}
  {34}},\ \bibinfo {pages} {3974} (\bibinfo {year} {1986})}\BibitemShut
  {NoStop}%
\bibitem [{\citenamefont {M\"uller}\ \emph {et~al.}(2012)\citenamefont
  {M\"uller}, \citenamefont {Stoklasa}, \citenamefont {Peuntinger},
  \citenamefont {Gabriel}, \citenamefont {\ifmmode \check{R}\else
  \v{R}\fi{}eh\'a\ifmmode~\check{c}\else \v{c}\fi{}ek}, \citenamefont {Hradil},
  \citenamefont {Klimov}, \citenamefont {Leuchs}, \citenamefont {Marquardt},\
  and\ \citenamefont {S\'anchez-Soto}}]{MSPG12}%
  \BibitemOpen
  \bibfield  {author} {\bibinfo {author} {\bibfnamefont {C.~R.}\ \bibnamefont
  {M\"uller}}, \bibinfo {author} {\bibfnamefont {B.}~\bibnamefont {Stoklasa}},
  \bibinfo {author} {\bibfnamefont {C.}~\bibnamefont {Peuntinger}}, \bibinfo
  {author} {\bibfnamefont {C.}~\bibnamefont {Gabriel}}, \bibinfo {author}
  {\bibfnamefont {J.}~\bibnamefont {\ifmmode \check{R}\else
  \v{R}\fi{}eh\'a\ifmmode~\check{c}\else \v{c}\fi{}ek}}, \bibinfo {author}
  {\bibfnamefont {Z.}~\bibnamefont {Hradil}}, \bibinfo {author} {\bibfnamefont
  {A.~B.}\ \bibnamefont {Klimov}}, \bibinfo {author} {\bibfnamefont
  {G.}~\bibnamefont {Leuchs}}, \bibinfo {author} {\bibfnamefont
  {C.}~\bibnamefont {Marquardt}}, \ and\ \bibinfo {author} {\bibfnamefont
  {L.~L.}\ \bibnamefont {S\'anchez-Soto}},\ }\href
  {http://stacks.iop.org/1367-2630/14/i=8/a=085002} {\bibfield  {journal}
  {\bibinfo  {journal} {New Journal of Physics}\ }\textbf {\bibinfo {volume}
  {14}},\ \bibinfo {pages} {085002} (\bibinfo {year} {2012})}\BibitemShut
  {NoStop}%
\bibitem [{\citenamefont {Bourgoin}\ \emph {et~al.}(2013)\citenamefont
  {Bourgoin}, \citenamefont {Meyer-Scott}, \citenamefont {Higgins},
  \citenamefont {Helou}, \citenamefont {Erven}, \citenamefont {H\"{u}bel},
  \citenamefont {Kumar}, \citenamefont {Hudson}, \citenamefont {D'Souza},
  \citenamefont {Girard}, \citenamefont {Laflamme},\ and\ \citenamefont
  {Jennewein}}]{BMHHE13}%
  \BibitemOpen
  \bibfield  {author} {\bibinfo {author} {\bibfnamefont {J.-P.}\ \bibnamefont
  {Bourgoin}}, \bibinfo {author} {\bibfnamefont {E.}~\bibnamefont
  {Meyer-Scott}}, \bibinfo {author} {\bibfnamefont {B.~L.}\ \bibnamefont
  {Higgins}}, \bibinfo {author} {\bibfnamefont {B.}~\bibnamefont {Helou}},
  \bibinfo {author} {\bibfnamefont {C.}~\bibnamefont {Erven}}, \bibinfo
  {author} {\bibfnamefont {H.}~\bibnamefont {H\"{u}bel}}, \bibinfo {author}
  {\bibfnamefont {B.}~\bibnamefont {Kumar}}, \bibinfo {author} {\bibfnamefont
  {D.}~\bibnamefont {Hudson}}, \bibinfo {author} {\bibfnamefont
  {I.}~\bibnamefont {D'Souza}}, \bibinfo {author} {\bibfnamefont
  {R.}~\bibnamefont {Girard}}, \bibinfo {author} {\bibfnamefont
  {R.}~\bibnamefont {Laflamme}}, \ and\ \bibinfo {author} {\bibfnamefont
  {T.}~\bibnamefont {Jennewein}},\ }\href
  {http://stacks.iop.org/1367-2630/15/i=2/a=023006} {\bibfield  {journal}
  {\bibinfo  {journal} {New Journal of Physics}\ }\textbf {\bibinfo {volume}
  {15}},\ \bibinfo {pages} {023006} (\bibinfo {year} {2013})}\BibitemShut
  {NoStop}%
\bibitem [{\citenamefont {Gabriel}\ \emph {et~al.}(2012)\citenamefont
  {Gabriel}, \citenamefont {Wittmann}, \citenamefont {Hacker}, \citenamefont
  {Mauerer}, \citenamefont {Huntington}, \citenamefont {Sabuncu}, \citenamefont
  {Marquardt},\ and\ \citenamefont {Leuchs}}]{Gabriel12}%
  \BibitemOpen
  \bibfield  {author} {\bibinfo {author} {\bibfnamefont {C.}~\bibnamefont
  {Gabriel}}, \bibinfo {author} {\bibfnamefont {C.}~\bibnamefont {Wittmann}},
  \bibinfo {author} {\bibfnamefont {B.}~\bibnamefont {Hacker}}, \bibinfo
  {author} {\bibfnamefont {W.}~\bibnamefont {Mauerer}}, \bibinfo {author}
  {\bibfnamefont {E.}~\bibnamefont {Huntington}}, \bibinfo {author}
  {\bibfnamefont {M.}~\bibnamefont {Sabuncu}}, \bibinfo {author} {\bibfnamefont
  {C.}~\bibnamefont {Marquardt}}, \ and\ \bibinfo {author} {\bibfnamefont
  {G.}~\bibnamefont {Leuchs}},\ }in\ \href
  {http://www.osapublishing.org/abstract.cfm?URI=CLEO_SI-2012-JW4A.103} {\emph
  {\bibinfo {booktitle} {Conference on Lasers and Electro-Optics 2012}}}\
  (\bibinfo  {publisher} {Optical Society of America},\ \bibinfo {year}
  {2012})\ p.\ \bibinfo {pages} {JW4A.103}\BibitemShut {NoStop}%
\bibitem [{\citenamefont {Ast}\ \emph {et~al.}(2013)\citenamefont {Ast},
  \citenamefont {Mehmet},\ and\ \citenamefont {Schnabel}}]{Ast2013}%
  \BibitemOpen
  \bibfield  {author} {\bibinfo {author} {\bibfnamefont {S.}~\bibnamefont
  {Ast}}, \bibinfo {author} {\bibfnamefont {M.}~\bibnamefont {Mehmet}}, \ and\
  \bibinfo {author} {\bibfnamefont {R.}~\bibnamefont {Schnabel}},\ }\href
  {\doibase 10.1364/OE.21.013572} {\bibfield  {journal} {\bibinfo  {journal}
  {Opt. Express}\ }\textbf {\bibinfo {volume} {21}},\ \bibinfo {pages} {13572}
  (\bibinfo {year} {2013})}\BibitemShut {NoStop}%
\end{thebibliography}%

\section*{Supplemental Material}

\subsection{\textit{A posteriori} probabilities for non-discretized homodyne detection}
\label{aposHD}
The \textit{a posteriori} probability for ideal, i.e. non-discretized, 
homodyne detection is derived from the statistical distribution of the 
measured quadrature variance 
$\sigma_N^2 = \sum_{k=1}^{N} x_k^2/N$, 
where $x_k$ are the continuous homodyne samples. 
The observed quadrature variance follows a $\chi^2_N$ distribution, 
which is scaled such that the expectation value coincides with the 
mean quadrature variance of the detected state.

Given the observation of the quadrature variance $\sigma_N^2$, 
the \textit{a posteriori} probability for the hypothesis 
$h\in\{\mathrm{coh}, \mathrm{sqz}\}$ is 
\begin{equation}
P_{H\left|Y\right.}\left(h\,|\,\sigma_N^2\right) = \frac{P_{Y\left|H\right.}\left(\sigma_N^2\left|\,h\right.\right)}{P_{Y\left|H\right.}\left(\sigma_N^2\left|\,\mathrm{coh}\right.\right)+P_{Y\left|H\right.}\left(\sigma_N^2\left|\,\mathrm{sqz}\right.\right)},
\label{eq:IdeaHDaposteriori}
\end{equation}

The average \textit{a posteriori} probability is obtained 
via integration over the scaled $\chi^2_N$ probability distribution 
for observing the quadrature variance $\sigma_N^2$ 
\begin{equation}
\langle P_{H} \rangle = \frac{1}{2}\sum_{h} \int_{0}^{\infty} P_{Y\left|H\right.}\left(\sigma_N^2\left|\,h\right.\right) \cdot P_{H\left|Y\right.}\left(h\,\left|\,\sigma_N^2\right.\right) \,\mathrm{d}\sigma_N^2
\label{eq:APOS_FullHD}
\end{equation}

In the state discrimination scenario, the state with 
the higher \textit{a posteriori} probability 
$\langle P_{H}(h) \rangle>1/2$ is hypothesized.

\begin{equation}
P_{H\left|Y\right.}\left(\mathrm{sqz}\left|\sigma_N^2\right.\right)
\begin{array}{c}
H=\mathrm{sqz}\\
>\\
<\\
H=\mathrm{coh}
\end{array}
P_{H\left|Y\right.}\left(\mathrm{coh}\left|\sigma_N^2\right.\right).
\end{equation}

\subsection{Discontinuous optimal displacement amplitude in maximizing the success probability}
\label{Discontinuity}

\begin{figure}[tb]%
\includegraphics[width=.70\columnwidth]{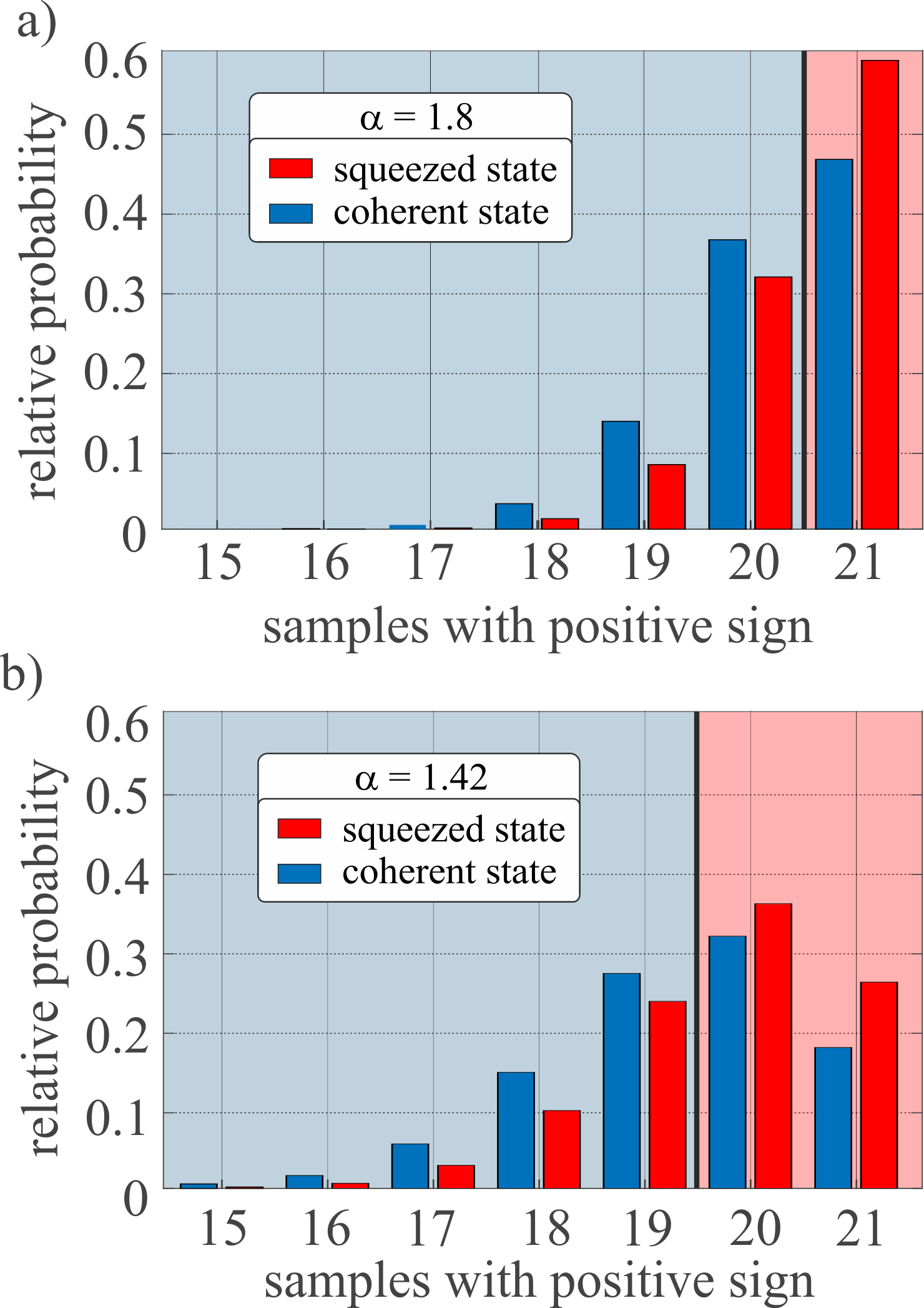}
\caption{Comparison of the statistical distributions for 
$N = 21$ samples and for different displacement amplitudes.
a) $\alpha=1.8$: opt. displacement just before the discontinuity at $N=20$.
a) $\alpha=1.42$: opt. displacement just after the discontinuity at $N=21$.}%
\label{fig:DiscreteAlpha}%
\end{figure}

The Binomial probability density functions (see Eq.(\ref{eq:bino_dist})) 
for the projection of 21 copies of the coherent state and the squeezed state 
onto the positive quadrature semi-axis $\Pi_{+}$ are shown for different 
displacement amplitudes $\alpha$ in Fig.~\ref{fig:DiscreteAlpha}. 
The blue-shaded areas indicates the cumulative quadrature-parity outcomes that 
lead to the hypothesis for the coherent state while outcomes in the red-shaded 
area are identified as the squeezed state. 
The distribution in Fig.~\ref{fig:MultiCopyCombined}(a) is obtained with the 
displacement amplitude maximizing the success probability for $N=20$ copies, 
$\alpha=1.8$, i.e. just before the discontinuity. 
In this configuration, the measurement hypothesizes the squeezed state only if all 
($k = 21$) detected states are projected onto a positive quadrature value. 
Fig.~\ref{fig:MultiCopyCombined}(b) shows the distribution at the actual 
optimal displacement amplitude $\alpha = 1.42$.
The state is identified as the squeezed state if $k=20$ or $k=21$ detections 
had a positive quadrature projection. 
The discontinuities in the optimized displacement curve result from discrete 
shifts in the decision threshold combined with the maximization of the 
likelihoods of the coherent and squeezed state distributions on their 
respective identification domains. 
For a large number of copies $N$ the range of possible outcomes approaches 
a quasi-continuous Gaussian distribution and the optimized displacement 
approaches an asymptotically optimal value.

\subsection{Required samples size for squeezing detection in LEO satellite links}
\label{datasize}

\begin{figure}[]%
\includegraphics[width=0.85\columnwidth]{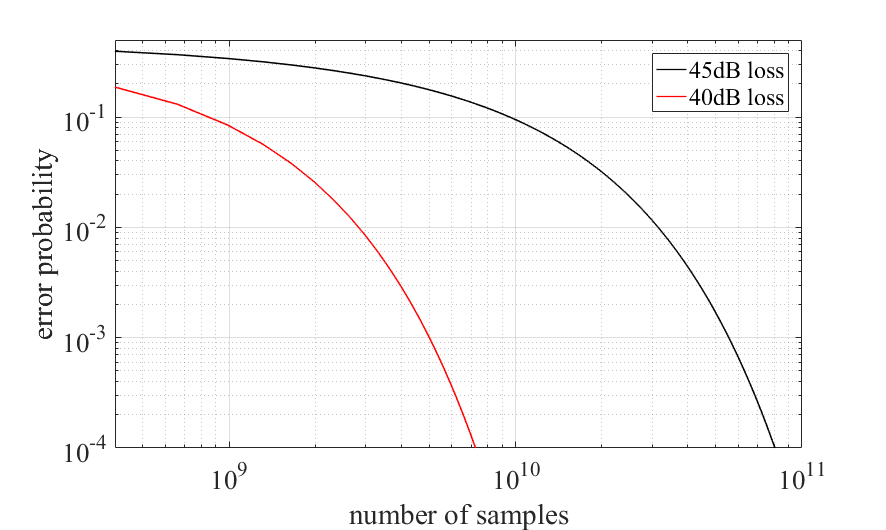}
\caption{Average minimum error probability as a function of 
the number of measured samples for the detection of a beam of light 
initially squeezed 6\,dB below the shot noise level at the output 
of a lossy channel of (a) 40\,dB loss, and (b) 45\,dB loss.}
\label{Chernoffreplacement}%
\end{figure}

Fig.~\ref{Chernoffreplacement} depicts the error probability for 
the unbiased detection of squeezed states (6\,dB, $r = 0.69$) 
emerging from a lossy channel of 40\,dB and 45\,dB. 
On an absolute scale,the error probability remains nearly constant 
up to a certain number of detected samples, but thereafter drops 
steeply with increasing number of samples.
The plot shows that squeezing can be detected with an average 
error probability no greater than $10^{-2}$ by measuring 
$3\cdot10^9$ (40\,dB) samples and $3\cdot10^{10}$ (45\,dB) samples, respectively. 

\end{document}